\documentclass[twocolumn, Journal]{IEEEtran}
\IEEEoverridecommandlockouts
\usepackage{bm}
\usepackage[english]{babel}
\usepackage{comment}
\usepackage{float}
\usepackage{subfig}
\usepackage{color}
\usepackage{graphicx}
\usepackage{amsmath}
\usepackage{amssymb}
\usepackage{algorithm}
\usepackage{algorithmic}
\usepackage{amsmath}
\usepackage{multirow}
\usepackage{booktabs}
\usepackage{array}
\usepackage{amsthm}
\usepackage{stfloats}
\usepackage{caption}
\usepackage{cuted}%%stripsep-3pt
\usepackage{setspace}
\usepackage{booktabs}   
\usepackage{multirow}    
\usepackage{siunitx}    
\usepackage[table]{xcolor}
\usepackage{caption}
\allowdisplaybreaks
\usepackage{lastpage}

\newcommand{\be}{\begin{equation}}
\newcommand{\ee}{\end{equation}}
\newcommand{\bea}{\begin{eqnarray}}
\newcommand{\eea}{\end{eqnarray}}
\newcommand{\ba}{\begin{array}}
\newcommand{\ea}{\end{array}}

\usepackage[table]{xcolor}
\usepackage{booktabs}
\usepackage{pifont}
\usepackage{graphicx}
 
\newcommand{\cmark}{\textcolor{green!60!black}{\ding{51}}}  %
\definecolor{lightgray}{gray}{0.9}
\definecolor{lightergray}{gray}{0.95}
\definecolor{lightblue}{RGB}{220,230,241}
\title{Graph Learning for  Cooperative Cell-Free ISAC Systems: From Optimization to Estimation
\thanks{Part of this paper will be presented at the IEEE Global Communication Conference (GLOBECOM 2025) \cite{add_ref1}.}
\thanks{P. Jiang and M. Li are with the School of Information and Communication Engineering, Dalian University of Technology, Dalian 116024, China (e-mail: pengjiang@mail.dlut.edu.cn; mli@dlut.edu.cn).}
\thanks{R. Liu is with the Department of Electrical Engineering and Computer Science, University of California Irvine, CA 92697, USA (e-mail: rangl2@uci.edu).}
\thanks{Q. Liu is with the School of Computer Science and Technology, Dalian University of Technology, Dalian 116024, China (e-mail: qianliu@dlut.edu.cn).}
}
\author{Peng Jiang, Ming Li,~\IEEEmembership{Senior Member,~IEEE}, Rang Liu,~\IEEEmembership{Member,~IEEE}, and Qian Liu,~\IEEEmembership{Member,~IEEE}}

\begin{document}
\maketitle
 \thispagestyle{empty}
\begin{abstract}
Cell-free integrated sensing and communication (ISAC) systems have emerged as a promising paradigm for sixth-generation (6G) networks, enabling simultaneous high-rate data transmission and high-precision radar sensing through cooperative distributed access points (APs). Fully exploiting these capabilities requires a unified design that bridges system-level optimization with multi-target parameter estimation. This paper proposes an end-to-end graph learning approach to close this gap, modeling the entire cell-free ISAC network as a heterogeneous graph to jointly design the AP mode selection, user association, precoding, and echo signal processing for multi-target position and velocity estimation. In particular, we propose two novel heterogeneous graph learning frameworks: a dynamic graph learning framework and a lightweight mirror-based graph attention network (mirror-GAT) framework. The dynamic graph learning framework employs structural and temporal attention mechanisms integrated with a three-dimensional convolutional neural network (3D-CNN), enabling superior performance and robustness in cell-free ISAC environments. Conversely, the mirror-GAT framework significantly reduces computational complexity and signaling overhead through a bi-level iterative structure with share adjacency. Simulation results validate that both proposed graph-learning-based frameworks achieve significant improvements in multi-target position and velocity estimation accuracy compared to conventional heuristic and optimization-based designs. Particularly, the mirror-GAT framework demonstrates substantial reductions in computational time and signaling overhead, underscoring its suitability for practical deployments.
\end{abstract}
\maketitle
 \pagestyle{empty}
%\vspace{-0.4 cm}
\section{Introduction}

Integrated sensing and communication (ISAC) has emerged as a pivotal technology for sixth‑generation (6G) wireless systems \cite{ITU}-\cite{Liu WCM 2023}. Unlike conventional single‑node ISAC architectures, a networked multi‑node cooperative ISAC paradigm leverages geographically distributed nodes to jointly perform communication and sensing. Such cooperation can substantially enhance spatial coverage and increase sensing accuracy through multi‑perspective observations.

\textcolor{black}{Among various network architectures, cell‑free massive multiple‑input multiple‑output (MIMO) has recently attracted significant attention as a promising foundation for cooperative ISAC. In a cell‑free system, a large number of distributed access points (APs) collaboratively serve multiple users without predefined cell boundaries. This dense and flexible topology inherently supports cooperative ISAC by providing high spatial diversity \cite{CF_ISAC}, effective interference management, and uniform user‑centric coverage \cite{ISAC_ZD}. These characteristics make the cell‑free architecture particularly suitable for simultaneously delivering high‑quality communication services and precise radar sensing \cite{add_ref2}.}

From the communication perspective, cell‑free MIMO mitigates cell‑edge degradation and shadowing by enabling each user to be jointly served by multiple nearby APs. This user‑centric cooperation enhances spectral efficiency and ensures uniformly high data rates across the entire coverage area \cite{ZhangVTM2021}. However, in practical large‑scale deployments, full connectivity between all users and APs is infeasible due to power, bandwidth, and complexity constraints. Therefore, adaptive user association and precoding strategies are crucial to fully exploit the spatial gains of cell‑free ISAC systems \cite{EmSP2021}.

\textcolor{black}{From the sensing perspective, the cell-free architecture inherently supports multi-static radar sensing by enabling the simultaneous illumination and observation of targets from multiple spatial angles \cite{add_ref4}. Such multi-perspective radar configurations effectively mitigate line-of-sight (LoS) blockage, reduce interference and clutter effects, and significantly enhance both target detection reliability and parameter estimation accuracy \cite{WCNC2025}. Meanwhile, by spatially separating the transmitter and receiver, the cell-free system eliminates the need for in-band full-duplex operation and complex self-interference cancellation \cite{XJSTSP2025}. }

\textcolor{black}{
Despite these advantages, coordinating the waveforms of multiple transmitters while maintaining mutual orthogonality remains a significant challenge in cooperative ISAC systems. Traditional multiplexing methods, such as time-division multiplexing (TDM) and frequency-division multiplexing (FDM), inherently face trade-offs between range and Doppler estimation performance. Alternatively, dynamically allocating distinct time-frequency resources to each AP enables simultaneous high-quality range and Doppler measurements, but at the expense of severe spectral inefficiency and increased scheduling complexity \cite{add_LPS_TSP}. In contrast, spatial multiplexing can effectively leverage carefully designed precoders and pseudo-orthogonal Zadoff-Chu (ZC) sequences, facilitating concurrent multi-AP transmissions with manageable interference, high orthogonality, and superior spectral efficiency. }

\textcolor{black}{To fully unlock the cooperative potential of cell-free ISAC networks, a unified optimization-estimation framework is required. Such a framework should jointly integrate AP operating mode (transmit or receive) selection, adaptive user association, precoding strategies to enhance target illumination and suppress clutter, as well as robust multi-target parameter estimation and network-level fusion algorithms \cite{add_ref3}.} While prior studies \cite{cellfreeisac}-\cite{Glob2023WS} have individually addressed certain aspects of this integrated problem, none have simultaneously considered all tightly coupled dimensions within a unified framework. For instance, existing works on joint AP mode selection and precoding \cite{cellfreeisac}-\cite{HYMTWC2024} typically neglect cooperative information fusion, while user-centric association and beamforming studies \cite{TWC2022, ZJ2023} overlook adaptive mode configuration. Similarly, earlier research on cooperative fusion and precoding \cite{WZHICASSP, ACTWC2024} lacks flexibility in association or mode adaptation. Furthermore, most of these existing approaches rely heavily on prior knowledge of target parameters, such as  \textcolor{black}{radar cross-section (RCS)}, angle-of-arrival (AoA), or accurate channel state information (CSI) \cite{HJTAES2024}, assumptions that rarely hold in practical scenarios where targets are unknown and environmental conditions vary dynamically. Consequently, these fragmented and assumption-dependent solutions fail to fully exploit the cooperative benefits offered by cell-free ISAC architectures. This underscores the critical need for a comprehensive, unified optimization-estimation framework that adaptively configures AP roles, user associations, and precoding strategies based on real-time sensing feedback rather than assumed target parameters.

\begin{table}[!t]
\centering
\scriptsize
\caption{Summary of studies on cell-free ISAC systems.}
\begin{tabular}{lcccccc}
\rowcolor{lightblue}
\toprule
\textbf{Ref.}  & \textbf{ISAC} & \textbf{Unknown} & \textbf{User} & \textbf{AP} & \textbf{Precoding} & \textbf{Signal} \\
\rowcolor{lightblue}
\textbf{}  & \textbf{Scenario} & \textbf{Target} & \textbf{Assoc.} & \textbf{Sel.} & \textbf{Design} & \textbf{Fusion} \\
\midrule
\cite{cellfreeisac}     & \cmark &  &  & \cmark & \cmark &  \\
\rowcolor{lightergray}
\cite{LXTCom}          & \cmark & \cmark &  & \cmark & \cmark &  \\
\cite{HYMTWC2024}      & \cmark & \cmark &  & \cmark & \cmark & \\
\rowcolor{lightergray}
\cite{TWC2022}          &  &  & \cmark &  & \cmark &  \\
\cite{ZJ2023}           & \cmark &  & \cmark &  & \cmark &  \\
\rowcolor{lightergray}
\cite{WZHICASSP}        & \cmark & \cmark &  &  & \cmark  & \cmark \\
\cite{ACTWC2024}       & \cmark & \cmark &  &  & \cmark & \cmark \\
\rowcolor{lightergray}
\cite{DBTCNS2020}       &        & \cmark & &  &  & \cmark \\
\cite{fzyTVT2024}       & \cmark & \cmark &  &  &  & \cmark \\
\rowcolor{lightergray}
\cite{A01}             & \cmark &  &  &  & \cmark &  \\
\cite{WZQTVT2024}       & \cmark &  &  &  & & \cmark \\
\rowcolor{lightergray}
\cite{A02}             & \cmark & \cmark &  &  & \cmark &  \\
\cite{A03}             & \cmark & \cmark & &  & \cmark &  \\
\rowcolor{lightergray}
\cite{AATcom2025}     & \cmark &  & & & \cmark &  \\
\cite{Glob2023WS}       &  &  &  &  & \cmark &  \\
\rowcolor{lightgray}
\textbf{Prop.}     & \cmark & \cmark & \cmark & \cmark & \cmark & \cmark \\
\bottomrule
\end{tabular}%
\label{summary}
% \vspace{-0.2cm}
\end{table}

Graph neural networks (GNNs) have recently emerged as a promising approach for addressing these challenges, given their inherent ability to effectively represent complex spatial correlations and hierarchical interactions among distributed APs and users in cell-free ISAC systems \cite{Glob2023AA}-\cite{WZH1}. Specifically, attention-based heterogeneous GNNs have been developed for joint sensing–communication channel modeling, demonstrating substantial performance improvements over conventional methods such as null-space projection and deep neural network (DNN) approaches \cite{WZH1}. In addition, link-heterogeneous GNNs have proven effective for cooperative localization by adaptively weighting sensing and communication links through attention-driven message passing \cite{LHGNN}. However, existing graph-learning methods typically address isolated subproblems and rely heavily on complete prior knowledge, and thus lack an integrated framework that jointly optimizes network configurations and target parameter estimation, especially under realistic conditions with unknown target information.

Motivated by the effectiveness of GNNs in capturing spatial and topological interactions, we propose two novel heterogeneous graph-learning frameworks to address the joint challenges of network-level optimization and target parameter estimation in cooperative cell-free ISAC systems. The primary contributions of this paper are summarized as follows:

\begin{itemize}

\item We propose a unified cooperative cell-free ISAC architecture that integrates dynamic AP mode selection, adaptive user association, cooperative transmit precoding, and distributed multi-target parameter estimation and fusion at the receiver APs. Crucially, our approach does not rely on prior knowledge of target parameters, significantly enhancing its practical applicability.

\item We develop a dynamic heterogeneous graph-learning framework based on a \textcolor{black}{graph attention network (GAT)}, enhanced by structural and temporal attention mechanisms. This framework jointly optimizes AP mode selection, user association, precoding strategies, and multi-target position and velocity estimation through a tailored 3D convolutional neural network module, resulting in substantial improvements in both sensing accuracy and multi-user communication performance.

\item We introduce a lightweight mirror-GAT framework to address computational complexity and signaling overhead issues. This framework leverages a bi-level iterative structure with shared-edge connections and distributed inference, achieving near-optimal performance at significantly reduced computational cost and backhaul requirements. It is thus well-suited for scalable and practical cell-free ISAC implementations.

\item We demonstrate through extensive simulations that the proposed frameworks consistently outperform heuristic and traditional optimization-based benchmarks, achieving sub-meter positioning accuracy and sub-m/s velocity estimation. Specifically, the mirror-GAT framework attains comparable accuracy to the dynamic approach but with considerably lower inference complexity and backhaul overhead, confirming its practical effectiveness for large-scale deployments.

\end{itemize}

\textit{Notations}: Scalar variables are denoted by normal-face letters, while vectors and matrices are denoted by bold lowercase and bold uppercase letters, respectively. $|a|$ and $\|\mathbf{a}\|$ denote the absolute value of scalar $a$ and the norm of vector $\mathbf{a}$, respectively. $\mathbb{C}$ and $\mathbb{R}$ represent the sets of complex and real numbers, respectively. $(\cdot)^T$ and $(\cdot)^H$ denote the transpose and conjugate transpose, respectively. $\mathbf{a}^{(i)}$ denotes the $i$-th iterative solution of $\mathbf{a}$. The real and imaginary parts of $a$ are denoted by $\Re\{\cdot\}$ and $\Im\{\cdot\}$, respectively. $\odot$ denotes the Hadamard product. $\underset{A}{\textrm{Top}}\{\cdot\}$ returns the indices of the $A$ largest elements.

\section{System Model and Problem Formulation}
\label{SystemModel}
\begin{figure}[!t]
\centering
\includegraphics[width=3.3 in]{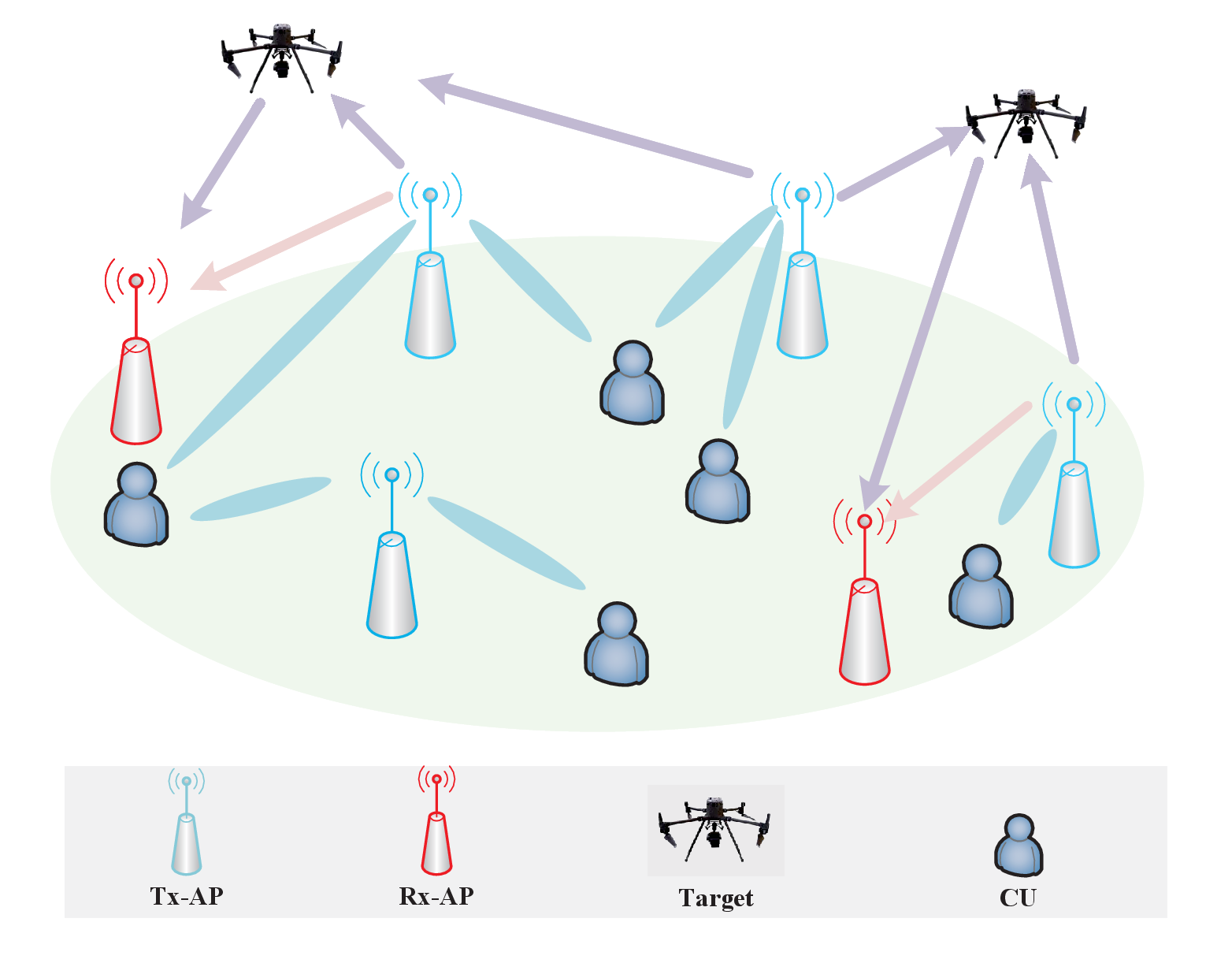}
\caption{Cell-free ISAC system.}
\label{system}
\vspace{-0.2 cm}
\end{figure}
We consider a cell-free ISAC system, as depicted in Fig.~\ref{system}. The network consists of $J$ distributed dual-function APs, each equipped with a uniform linear array (ULA) of $M$ antennas spaced at half-wavelength intervals. These APs cooperatively serve $K$ single-antenna communication users (CUs) while simultaneously sensing $Q$ potential targets. \textcolor{black}{Specifically, let $\mathbf{p}_q \in \mathbb{R}^2$ and $\mathbf{v}_q \in \mathbb{R}^2$ denote the position and velocity vectors of the $q$‑th target, respectively. The two elements of $\mathbf{p}_q$ and $\mathbf{v}_q$ correspond to the components along the $x$‑ and $y$‑axes, respectively.}

In this cell-free ISAC system, each AP can dynamically switch its operation mode between transmission (for downlink communication and target illumination) and dedicated sensing reception (for target estimation). This mode-switching flexibility allows the system to fully exploit available spatial degrees of freedom (DoFs), leverage multi-perspective observations, and avoid interference and clutter, thereby substantially enhancing overall system performance.
Let $T$ denote the number of APs operating in transmit mode, and $R$ the number operating in receive mode, such that $T + R = J$. The corresponding sets of AP indices are defined as $\mathcal{T}$ and $\mathcal{R}$, respectively. We introduce a binary mode selection vector $\bm{\kappa} = [\kappa_1,\kappa_2,\ldots,\kappa_J]^T \in \{0,1\}^J$, where $\kappa_j = 1$ if AP $j$ is in transmit mode and $\kappa_j = 0$ otherwise. To ensure proper system configuration, the mode selection must satisfy $\sum_{j=1}^J \kappa_j = T$, thereby guaranteeing that exactly $T$ APs operate in transmit mode.

In addition to AP operating mode selection, an effective AP-CU association is also important to fully harness the spatial diversity offered by both APs and CUs. This association, which assigns each transmit-mode AP (Tx-AP) specific CUs to serve, is crucial for efficient AP resource utilization, balanced load distribution across APs, and high-quality service for all users. To explicitly characterize this association, we define the binary association matrix $\bm{\Lambda}\in \{0,1\}^{J \times K}$, where $\Lambda_{j,k} = 1$ if and only if the $j$-th AP operates in transmit mode and serves the $k$-th CU; otherwise, $\Lambda_{j,k} = 0$. Specifically, we set $\Lambda_{j,k} = 0$ for all CUs whenever $j \in \mathcal{R}$, reflecting the fact that receive-mode APs (Rx-APs) do not serve any CUs. In addition, if each Tx-AP simultaneously serves at most $K_{\mathrm{u}}$ designated CUs, the constraint $\sum_{k=1}^K \Lambda_{j,k} \leq K_{\mathrm{u}}$ must be enforced.

\begin{table}[!t]
\centering
\footnotesize
\color{black}
\caption{System Notations.}
\label{notation}
\renewcommand{\arraystretch}{1.2}
\begin{tabular}{@{} l @{\hspace{3pt}} c @{\hspace{3pt}} !{\vrule width 1pt} @{\hspace{3pt}} l @{\hspace{3pt}} c @{}}
\hline
\textbf{Description} & \textbf{Notation} & \textbf{Description} & \textbf{Notation} \\
\hline
Carrier frequency & $f_c$ & Bandwidth & $B$ \\
No. of subcarriers & $N_{\rm s}$ & No. of OFDM syms. & $L$ \\
No. of APs & $J$ & No. of Tx-APs & $T$ \\
No. of Rx-APs & $R$ & No. of CUs & $K$ \\
No. of assoc. CUs & $K_{\rm u}$ & No. of targets & $Q$ \\
No. of clutters & $U$ & No. of antennas & $M$ \\
Transmit power & $P_j$ & Commun. noise & $\sigma_{\rm c}^2$ \\
Sensing noise& $\sigma_{\rm r}^2$ & Position of target-$q$ & $\mathbf{p}_q$ \\
No. of hidden layers & $\tau^{\rm tot}_1~\&~\tau^{\rm tot}_2$ & Hidden layer index & $\tau_1~\&~\tau_2$ \\
No. of snapshots & $N$ & No. of mirror iter. & $N_{\rm m}$ \\
Comp. Dim. & $\varrho$ & Agg. Dims. & \hspace{-0.3 cm} $D_{\rm CU}\&D_{\rm Rx}\&D_{\rm AP}$ \\
\hline
\end{tabular}
\vspace{-0.3 cm}
\end{table}

The considered cell-free ISAC system utilizes orthogonal frequency-division multiplexing (OFDM) waveforms for both communication and sensing functions. Specifically, each Tx-AP constructs OFDM frames consisting of $N_{\mathrm{s}}$ orthogonal subcarriers spanning $L$ OFDM symbols. The subcarrier spacing is denoted by $\Delta f = B/N_{\mathrm{s}}$, with $B$ being the total available bandwidth. Each OFDM symbol includes a cyclic prefix (CP) of duration $T_{\mathrm{p}}$, resulting in a total OFDM symbol duration $T_{\mathrm{sym}} = 1/\Delta f + T_{\mathrm{p}}$.

For the $i$-th subcarrier during the $l$-th OFDM symbol, let $\mathbf{s}_{\mathrm{c},i}[l]\triangleq[s_{\mathrm{c},i,1}, s_{\mathrm{c}, i,2},\ldots,s_{\mathrm{c}, i, K}]^T\in\mathbb{C}^{K\times1}$ denote the vector of modulated communication symbols for the $K$ CUs. Additionally,  for the $j$-th Tx-AP, let $\mathbf{s}_{\mathrm{r},j,i}[l] \in \mathbb{C}^{M\times 1}$ represent the sensing (radar probing) symbol vector generated from a ZC sequence. It is worth emphasizing that different Tx-APs employ distinct and quasi-orthogonal ZC sequences with coprime root indices. This strategy enables Rx-APs to reliably distinguish echo signals originating from different TX-APs, thereby facilitating multi-perspective observation and preventing ghost targets.

\textcolor{black}{To simultaneously multiplex communication symbols and radar sensing signals onto the same subcarrier-$i$, each Tx-AP $j \in \mathcal{T}$ employs precoding matrices specifically tailored for these dual functionalities. Let $\mathbf{W}_{\mathrm{c},j,i}\triangleq[\mathbf{w}_{\mathrm{c},j,i,1},\ldots,\mathbf{w}_{\mathrm{c},j,i,K}] \in \mathbb{C}^{M \times K}$ denote the transmit precoding matrix associated with communication signals, and let $\mathbf{W}_{\mathrm{r},j,i} \triangleq [\mathbf{w}_{\mathrm{r},j,i,1},\ldots,\mathbf{w}_{\mathrm{r},j,i,M}]\in \mathbb{C}^{M \times M}$ represent the transmit precoding matrix for radar sensing. Given that Rx-APs do not engage in transmit precoding, we conveniently set their corresponding precoding matrices to zero, i.e., $\mathbf{W}_{\mathrm{c},j,i}\triangleq \mathbf{0}^{M \times K}$ and $\mathbf{W}_{\mathrm{r},j,i}\triangleq \mathbf{0}^{M \times M}$ for each Rx-AP $j \in \mathcal{R}$. Consequently, the transmitted signal vector from the $j$-th Tx-AP on subcarrier-$i$ during the $l$-th OFDM symbol can be expressed as
\begin{align}
\mathbf{x}_{j,i}[l] = \kappa_{j}\mathbf{W}_{\mathrm{r},j,i}\mathbf{s}_{\mathrm{r},j,i}[l] + \kappa_{j}\sum_{k=1}^K\Lambda_{j,k}\mathbf{w}_{\mathrm{c},j,i,k} s_{\mathrm{c},i,k}[l],
\end{align}
where $\kappa_{j}$ indicates the operational mode of AP-$j$ (transmitting or not), and $\Lambda_{j,k}$ denotes the user association indicator between AP-$j$ and the $k$-th CU.
}

\subsection{Multi-User Communication Metric}
\textcolor{black}{For the communication function, the received signal at the $k$-th CU on the $i$-th subcarrier during the $l$-th time slot is represented as
\be
y_{\rm c, \it i,k}[l] = \sum_{j=1}^{J}\mathbf{h}_{j,i,k}^{H}\mathbf{x}_{j,i}[l] + n_{\rm c, \it i, k}[l],
\ee
where $\mathbf{h}_{j,i,k}\in\mathbb{C}^{M\times1}$ denotes the baseband channel from the $j$-th AP to the $k$-th CU on the $i$-th subcarrier, and $n_{\rm c, \it i, k}[l]\sim\mathcal{CN}(0,\sigma_{\rm c}^2)$ denotes the additive white Gaussian noise (AWGN).} The corresponding SINR for the $k$-th CU on the $i$-th subcarrier is given by
\be
\textrm{SINR}_{\rm c, \it i, k}=\frac{\big|\sum\limits_{j=1}\limits^{J}\kappa_{j}\Lambda_{j,k}\mathbf{h}_{j,i,k}^{H}\mathbf{w}_{{\rm c}, j,i,k}\big|^2}{\mu_{i,k}+\iota_{i,k}+\sigma_{\rm c}^2},
\ee
where $\mu_{i,k}$ and $\iota_{i,k}$ denote the multi-user interference and radar interference, respectively, and are defined as follows:
\begin{subequations}\begin{align}
\mu_{i,k}&=\sum\limits_{n=1,n\neq k}\limits^{K}\big|\sum\limits_{j=1}\limits^{J}\kappa_{j}\Lambda_{j,n}\mathbf{h}_{j,i,k}^{H}\mathbf{w}_{\rm c,\it j,i, n}\big|^2,   \\     \iota_{i,k}&=\sum\limits_{m=1}\limits^{M}\big|\sum\limits_{j=1}\limits^{J}\kappa_{j}\mathbf{h}_{j,i,k}^{H}\mathbf{w}_{\rm r, \it j,i, m}\big|^2.
\end{align}\end{subequations}
We employ the achievable sum-rate as the communication performance metric for the cell-free ISAC system, which can be calculated as
\begin{align}
\label{commun_metric}
R_\text{c} =\sum\limits_{k=1}^{K}\sum\limits_{i=1}^{N_s}\textrm{log}_2(1+\textrm{SINR}_{\rm c, \it i, k}).
\end{align}

\begin{figure*}[!t]
\begin{align}\label{eq:yr}
\mathbf{y}_{r,i}[l] & =\sum_{j=1}^{J} \sum_{q=1}^{Q} \beta_{j,r,q} \sqrt{\text{PL}(d_{r,q})\text{PL}(d_{j,q})}  \mathbf{a}_{\rm r} (\theta_{r,q})\mathbf{a}_{\rm t}^T (\theta_{j,q}) \mathbf{x}_{j,i}[l]e^{-\jmath 2 \pi i \Delta f \left(d_{r,q} + d_{j,q}\right)/c} e^{\jmath 2 \pi l T_\text{sym} f_{\rm d, \it j,r,q}} \nonumber \\
&\quad + \sum_{j=1}^{J} \sum_{u=1}^{U} \eta_{j,r,u} \sqrt{\text{PL}(d_{r,u})\text{PL}(d_{j,u})} \mathbf{a}_{\rm r} (\theta_{r,u})  \mathbf{a}_{\rm t}^T (\theta_{j,u}) \mathbf{x}_{j,i}[l]e^{-\jmath 2 \pi i \Delta f \left( d_{r,u} + d_{j,u} \right)/c} e^{\jmath 2 \pi l T_\text{sym}  f_{\rm d, \it j,r,u}} \nonumber \\
&\quad + \sum_{j=1}^{J} \sqrt{\text{PL}(d_{j,r})}\mathbf{G}_{j,r}\mathbf{x}_{j,i}[l]e^{-\jmath 2 \pi i \Delta f d_{j,r}/c}+\mathbf{z}_{r,i}[l],
\end{align}
\rule[-0pt]{18 cm}{0.05em}
\end{figure*}

\subsection{Radar Sensing Metric}
For the radar sensing function, the received signal at an Rx‑AP consists of target echoes, clutter, direct inter‑AP interference, and AWGN. Specifically, the frequency‑domain baseband echo received at the $r$‑th Rx‑AP is given by \eqref{eq:yr} at the top of this page. 
The first term in \eqref{eq:yr} corresponds to the target echoes, where $\beta_{j,r,q}$ denotes the radar cross section (RCS) of the $q$‑th target illuminated by Tx‑AP-$j$ and observed by Rx‑AP-$r$, satisfying $\mathbb{E}{|\beta_{j,r,q}|^2} = \zeta_q^2$. The parameters $d_{j,q}$ and $d_{r,q}$ represent the distances from the $q$‑th target to Tx‑AP-$j$ and Rx‑AP-$r$, respectively, associated with angles $\theta_{j,q}$ and $\theta_{r,q}$. 
\textcolor{black}{Moreover, $f_{\mathrm{d},j,r,q}$ denotes the Doppler frequency shift caused by the radial velocity component of the $q$‑th target relative to Rx‑AP-$r$, induced by signals transmitted from Tx‑AP-$j$.} The speed of light is $c$, and the distance‑dependent path loss is modeled by $\mathrm{PL}(d)$ under a LoS assumption. The array steering vector at angle $\theta$ is defined as $\mathbf{a}(\theta) \triangleq [1, e^{\jmath\pi\sin\theta}, \ldots, e^{\jmath(M-1)\pi\sin\theta}]^T \in \mathbb{C}^{M\times 1}$. The second term in \eqref{eq:yr} represents clutter interference, where $\eta_{j,r,u}$ is the RCS of the $u$‑th clutter scatterer, with $\mathbb{E}{|\eta_{j,r,u}|^2} = \zeta_u^2$. The corresponding parameters, including distances $d_{r,u}$ and $d_{j,u}$, angles $\theta_{r,u}$ and $\theta_{j,u}$, and Doppler frequency $f_{\mathrm{d},j,r,u}$, are defined analogously to those of the targets.
The third term describes direct AP‑to‑AP interference, characterized by the LoS channel matrix $\mathbf{G}_{j,r}$ between Tx‑AP-$j$ and Rx‑AP-$r$. Finally, $\mathbf{z}_{r,i}[l]$ denotes AWGN, modeled as a complex Gaussian random vector $\mathcal{CN}(\mathbf{0}, \sigma_{\mathrm{r}}^2 \mathbf{I}_M)$.

All received echo signals $\mathbf{y}_{r,i}[l],~\forall r,i,l$, are aggregated into a data tensor $\mathbf{Y}_{\mathrm{r}} \in \mathbb{C}^{R \times N_{\mathrm{s}} \times L \times M}$, encapsulating range, velocity, and angular information of the targets. Given the observed echo tensor $\mathbf{Y}_{\mathrm{r}}$ at Rx‑APs and the known sensing sequence tensor $\mathbf{S}_{\mathrm{r}}\in\mathbb{C}^{T\times M \times N_{\mathrm{s}} \times L}$, the joint estimation of the positions and velocities of the $Q$ targets can be formulated as the following mapping function:
\begin{equation}
\label{mapping}
\{\hat{\mathbf{p}}_q, \hat{\mathbf{v}}_q \}_{q=1}^{Q} \;=\; \mathcal{M}( \mathbf{Y}_{\mathrm{r}}, \mathbf{S}_{\mathrm r})\,.
\end{equation}
The estimation performance is evaluated using the root-mean-square error (RMSE) metrics, which quantify estimation accuracy by measuring the deviations between the estimated and true positions and velocities of the targets. These metrics are formally defined as follows:
\begin{eqnarray}
\label{sensing_metric1}
\mathrm{RMSE_p}  &=& \sqrt{\,\mathbb{E}\Bigg[\frac{1}{Q}\sum_{q=1}^{Q} \big\|\,\mathbf{p}_{q} - \hat{\mathbf{p}}_{q}\,\big\|^2\Bigg]}, \\
   \label{sensing_metric2}
\mathrm{RMSE_v}  &=& \sqrt{\,\mathbb{E}\Bigg[\frac{1}{Q}\sum_{q=1}^{Q} \big\|\,\mathbf{v}_{q} - \hat{\mathbf{v}}_{q}\,\big\|^2\Bigg]},
\end{eqnarray}
where $\hat{\mathbf{p}}_{q}$ and  $\hat{\mathbf{v}}_{q}$ denote the estimated position and velocity, respectively.

\subsection{Problem Formulation}
Our objective is to jointly design the AP mode selection vector $\bm{\kappa}$, the user association matrix $\bm{\Lambda}$, the transmit precoding matrices $\mathbf{W}_{j,i} \triangleq [\mathbf{W}_{\mathrm{r},j,i}, \mathbf{W}_{\mathrm{c},j,i}] \in \mathbb{C}^{M \times (M+K)}$, $\forall j, i$, and the estimation algorithm $\mathcal{M}$, aiming to minimize the radar sensing errors while ensuring satisfactory multi-user communication throughput. In particular, we seek to minimize the weighted sum of position and velocity estimation RMSEs defined in \eqref{sensing_metric1} and \eqref{sensing_metric2}, subject to constraints on the multi-user achievable sum-rate $R_{\mathrm{c}}$ (given in \eqref{commun_metric}), transmit power, and AP-user association.
 \textcolor{black}{To ensure fairness in power allocation and reduce scheduling overhead in dynamic networks, we assume fixed numbers of Tx-APs, Rx-APs, and a fixed maximum number of associated users per Tx-AP. Consequently, the primary focus of the optimization is on assigning appropriate AP operating modes and user associations, which aligns with realistic system constraints and simplifies the problem formulation.} Mathematically, the joint optimization problem is expressed as:
\begin{subequations}\label{initial}
\begin{align}
&\underset{ \bm{\kappa},\,\bm{\Lambda}, \mathbf{W}_{j,i}, \mathcal{M} }{\min}~  \quad  \omega\mathrm{RMSE_p} + (1-\omega)\mathrm{RMSE_v} \label{obj_initial}\\
&\text{s.t.}~  R_{\mathrm{c}}  \geq\gamma,\label{commun_const}\\
&~~~~ \sum_{i=1}^{N_s} \big\| \mathbf{W}_{j,i} \big\|_{F}^{2} \;\le\; P_j,   \forall\, j \in \mathcal{T}, \label{power_const}\\
& ~~~~ \sum_{j=1}^J\kappa_j =  T,   \kappa_{j} \in \{0,1\}, \forall\, j \in \mathcal{J}, \label{kappa_binary} \\
& ~~~~ \sum_{k=1}^K \Lambda_{j,k} \leq K_{\mathrm{u}},  \Lambda_{j,k} \in \{0,1\},  \forall\, j \in \mathcal{J}, \forall\, k \in \mathcal{K}, \label{assoc_count}
\end{align}
\end{subequations}
where $0 \leq \omega \leq 1$ denotes the task weight factor, $\gamma$ represents the required minimum achievable sum-rate, and $P_j$ is the transmit power budget at Tx-AP-$j$. 

    \begin{figure*}[!t]
\centering
\includegraphics[width=6.3 in]{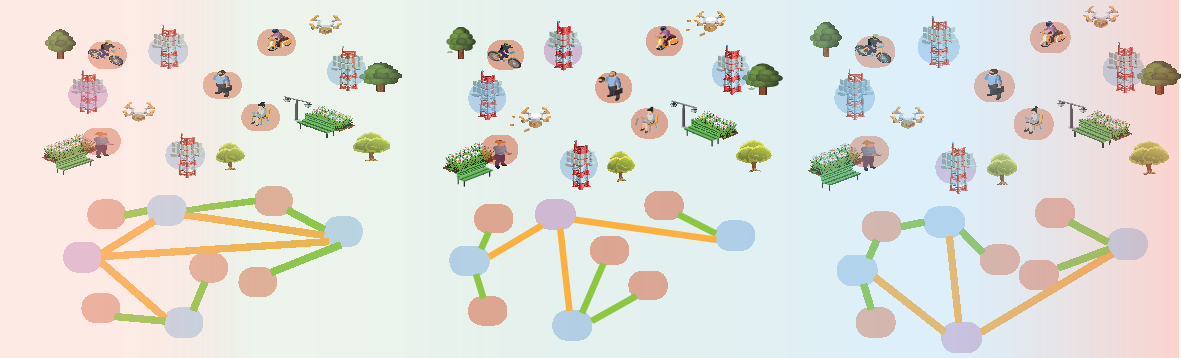}
\caption{Cell-free ISAC scenario and its corresponding heterogeneous graph structure.}
\label{scenario}
\end{figure*}

This joint optimization problem involves discrete decisions (AP mode selection and user association), continuous-valued precoding matrices, and an implicit estimation function, thereby posing significant technical challenges. First, the mixed-integer and non-convex structure of the problem precludes direct application of standard convex optimization methods. Second, the absence of prior knowledge regarding target parameters prevents explicit derivation of classical analytical bounds, such as the Cramér-Rao bound (CRB), which makes it difficult to directly quantify estimation accuracy in relation to system parameters. Consequently, conventional optimization-based approaches cannot effectively determine optimal AP configurations and precoding strategies. Third, the multi-target estimation function $\mathcal{M}$ in \eqref{mapping} inherently requires centralized fusion of signals received at spatially distributed APs. Such centralized fusion processes introduce considerable computational complexity and communication overhead, rendering traditional estimation techniques impractical, particularly given the high-dimensional data collected by multiple dispersed APs.

To address these challenges, we propose two novel graph-learning-based methodologies specifically designed for cooperative cell-free ISAC systems. By leveraging advanced graph neural network (GNN) architectures, our proposed frameworks efficiently handle the complex discrete-continuous optimization and centralized multi-target estimation tasks, thus fully exploiting the cooperative gains inherent in cell-free ISAC architectures.

\section{Proposed Dynamic Graph Learning Framework} \label{A}

In this section, we first define the heterogeneous dynamic graph structure and its corresponding cell-free ISAC system scenario, highlighting its deployment advantages. Subsequently, we propose a novel dynamic graph learning framework based on heterogeneous GNNs. Finally, we present the training strategy adopted for the proposed framework.

\subsection{Graph Definition}
\label{Graph_define}
As depicted in Fig.~\ref{scenario}, the considered cooperative cell-free ISAC system can be modeled as a heterogeneous graph consisting of multiple node and edge types. Specifically, each Tx-AP, Rx-AP, and CU corresponds to a unique node type, illustrated as blue, purple, and red nodes, respectively. Communication links are represented as edges connecting Tx-AP nodes to CU nodes (shown in green), while sensing links correspond to edges connecting Tx-AP nodes to Rx-AP nodes (shown in yellow). It is important to note that sensing edges reflect not only direct interference channels between Tx-APs and Rx-APs but also echo signals from potential targets. However, as prior information about target locations is initially unavailable, these echo signals are not explicitly modeled as edges in the initial graph representation; instead, they are incorporated as asynchronous compensatory features.
This heterogeneous graph formulation intuitively captures complex connectivity patterns within the cell-free ISAC system, including AP operating modes and AP-user associations. Moreover, the graphical structure effectively models multi-user and radar interference effects as propagation characteristics along edges. By explicitly embedding the spatial and topological information of cell-free networks into a GNN framework, the proposed approach effectively leverages network structures to facilitate joint optimization of communication and sensing functionalities.

\begin{figure*}[!t]
\centering
\includegraphics[width=5.5 in]{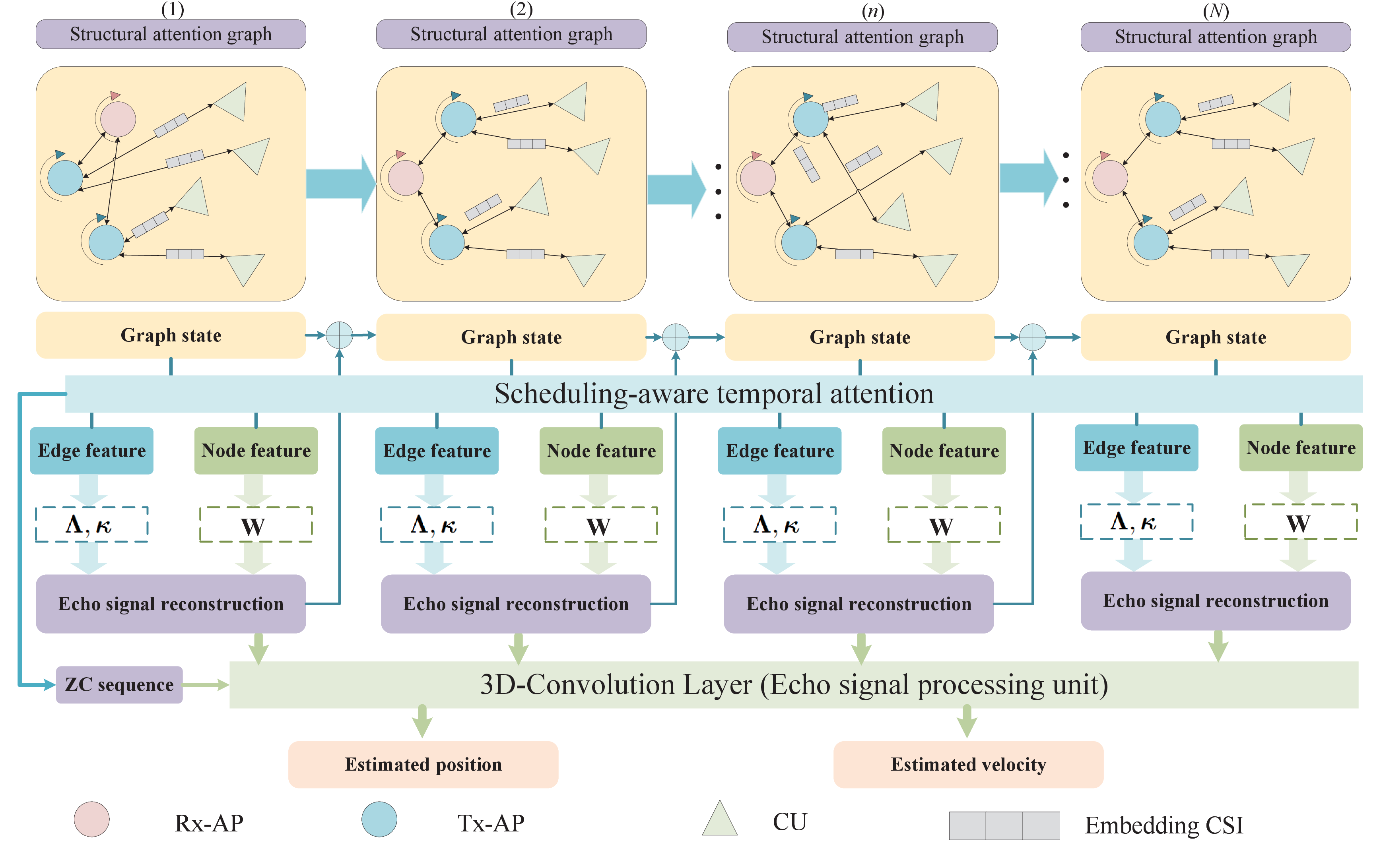}
\caption{Proposed dynamic graph learning framework.}\vspace{-0.2 cm}
\label{network}
\end{figure*}

However, due to the lack of prior information about potential targets, employing a static topology graph significantly restricts the capability to achieve globally optimal designs. Specifically, fixed AP modes and user associations, corresponding to an unchanging graph topology, fail to fully exploit the spatial DoFs inherent in cell-free ISAC deployments. Additionally, precoding matrices derived from static inference processes inherently lack adaptive power allocation, which degrades estimation accuracy for targets with unknown parameters.

To address these limitations and enable adaptive exploratory sensing of unknown targets, we propose a novel \textit{dynamic} heterogeneous graph-based framework comprising multiple temporal snapshots with evolving graph topologies \cite{graph}. In this dynamic framework, each temporal snapshot represents an instantaneous graph featuring a specific topology configuration. \textcolor{black}{The graph dynamically alters its edge connectivity from one snapshot to another. Formally, the cell-free ISAC network at the $n$-th snapshot is represented as a heterogeneous graph $\mathcal{G}^{(n)}=(\mathcal{V}^{(n)},\mathcal{E}^{(n)})$, where $\mathcal{V}^{(n)}$ and $\mathcal{E}^{(n)}$ denote the snapshot-specific node set and edge set, respectively. The node set comprises three distinct node types: Tx-AP nodes, Rx-AP nodes, and CU nodes, explicitly denoted as $\mathcal{V}^{(n)}=\{\mathcal{V}^{(n)}_{\mathrm{Tx}},\mathcal{V}^{(n)}_{\mathrm{Rx}},\mathcal{V}_{\mathrm{CU}}\}$. Within each snapshot, dynamic edges connect Tx-AP nodes to CU nodes and Tx-AP nodes to Rx-AP nodes, capturing instantaneous communication and sensing interactions, respectively.}

\textcolor{black}{Notably, adjacency relationships in the GNN do not directly correspond to physical associations. Instead, Tx-AP nodes maintain potential connections with all Rx-AP and CU nodes. Actual associations and AP operating modes are determined by the discrete optimization variables $\bm{\Lambda}$ and $\bm{\kappa}$. To align these continuous attention-based relationships with the discrete network structure, we introduce trainable edge information vectors with attention weights, effectively transforming continuous attention maps into topological configurations consistent with the physical cell-free ISAC architecture.}

\subsection{Graph Initialization}
\label{Dynamic_label_initial}

After defining the dynamic heterogeneous graph, we embed prior information from the cell-free ISAC system into the initial graph state to effectively address optimization problem \eqref{initial}. 
Specifically, by adopting the embedding CSI\footnote{\textcolor{black}{This work primarily addresses AP-user association, precoding design, and sensing fusion tasks. We assume CSI can be obtained via standard channel estimation techniques for cell-free MIMO systems \cite{EMCE2020}.}} of the communication link, the feature associated with an edge between the $t$-th Tx-AP and the $k$-th CU is defined as
\begin{align}\label{eq:htk}
\tilde{\mathbf{h}}_{t,k} \! = \!  \bigl[\mathrm{vec}^T(\Re\{\mathbf{h}_{t,:,k}\}), \mathrm{vec}^T(\Im\{\mathbf{h}_{t,:,k}\})\bigr] \! \in\mathbb{R}^{1\times 2N_{\mathrm{s}}M}.
\end{align}
Similarly, using embedding AP-to-AP interference, the feature associated with an edge between the $t$-th Tx-AP and the $r$-th Rx-AP is given by
\begin{align}\label{eq:Gtr}
\tilde{\mathbf{g}}_{t,r}=\bigl[\mathrm{vec}^T(\Re\{\mathbf{G}_{t,r}\}), \mathrm{vec}^T(\Im\{\mathbf{G}_{t,r}\})\bigr]\in\mathbb{R}^{1\times 2M^2}.
\end{align}

\textcolor{black}{Then, each snapshot graph is processed by a GNN with $\tau^{\mathrm{tot}}_1$ hidden layers.}  \textcolor{black}{The initial hidden states of Tx-AP, Rx-AP, and CU nodes are constructed by concatenating their corresponding edge features:
\begin{align}
\mathbf{s}^{(0)}_{\mathrm{Tx}}(t) &= \big[\, \tilde{\mathbf{h}}_{t,1},\, \tilde{\mathbf{h}}_{t,2},\, \ldots,\, \tilde{\mathbf{h}}_{t,K},\, \tilde{\mathbf{g}}_{t,1},\, \ldots,\, \tilde{\mathbf{g}}_{t,R} \,\big], \label{Tx-AP_feature}\\
\mathbf{s}^{(0)}_{\mathrm{Rx}}(r) &= \big[\, \tilde{\mathbf{h}}_{r,1},\, \tilde{\mathbf{h}}_{r,2},\, \ldots,\, \tilde{\mathbf{h}}_{r,K},\, \tilde{\mathbf{g}}_{1,r},\, \ldots,\, \tilde{\mathbf{g}}_{T,r} \,\big], \label{Rx-AP_feature}\\
\mathbf{s}^{(0)}_{\mathrm{CU}}(k) &= \big[\, \tilde{\mathbf{h}}_{1,k},\, \tilde{\mathbf{h}}_{2,k},\, \ldots,\, \tilde{\mathbf{h}}_{T,k} \,\big]. \label{user_feature}
\end{align}}

\subsection{Graph Module}
\label{Graph_module}
After graph initialization, we partition the optimization problem \eqref{initial} into three interconnected modules: user association, AP mode selection, and signal processing for target position and velocity estimation, as illustrated in Fig.~\ref{network}. \textcolor{black}{Specifically, a structural attention module first learns the user association matrix $\bm{\Lambda}$. 
Then, a temporal attention module exploits the received echo signals to dynamically optimize the AP mode selection vector $\bm{\kappa}$ for the next snapshot. 
Based on the Tx-AP node features obtained at the final snapshot, the optimal precoding matrix $\mathbf{W}_{j,i}$ is derived for transmission. 
Finally, a three-dimensional convolutional neural network (3D-CNN) module \cite{CNN}, trained offline using representative echo data, processes the received signals to extract essential spatial-frequency-temporal features across the antenna, subcarrier, and symbol dimensions. 
The 3D-CNN operates as a pre-trained feature extractor during initialization rather than as part of the dynamic graph optimization loop, ensuring a consistent and causal algorithmic flow.}

\subsubsection{Structural Attention Module}
In each snapshot, the hidden states of the Tx-AP nodes are iteratively updated using a masked structural attention mechanism that integrates attention scoring, message passing, and aggregation:
\begin{equation}
\mathbf{s}_{\mathrm{Tx}}^{(\tau_1+1,n)}(t)=f_{\mathrm{agg}}\Big(f_{\mathrm{att}}\big(\mathbf{s}_{\mathrm{Tx-AP}}^{(\tau_1,n)}(t)\big)\odot\underset{\xi\in\mathcal{N}^{(n)}(t)}{f_{\mathrm{mp}}}\big(\mathbf{s}^{(\tau_1,n)}(\xi)\big)\Big),
\end{equation}
where $f_{\mathrm{agg}}(\cdot)$ aggregates node features, $f_{\mathrm{att}}(\cdot)$ computes attention scores, and $f_{\mathrm{mp}}(\cdot)$ conducts message passing from neighboring nodes. \textcolor{black}{$\tau_1$ denotes the hidden layer index for the subgraph of the dynamic graph learning framework in each snapshot.} \textcolor{black}{Here, $\mathcal{N}^{(n)}(t)$ denotes the neighbor set connected to the $t$-th Tx-AP node in the $n$-th snapshot.} For clarity, the snapshot index $n$ is omitted in subsequent equations of structural attention module, as the graph topology remains unchanged within a snapshot.

For efficient utilization of power and bandwidth resources, each Tx-AP must selectively associate with appropriate CUs. \textcolor{black}{Within the structural attention module, the user association matrix $\bm{\Lambda}$ is obtained via a masked attention mechanism. For each Tx-AP, the effective feature scores of all neighboring nodes are evaluated, and the Top-$K_{\rm u}$ function selects the strongest $K_{\rm u}$ communication links for feature propagation according to their normalized attention probabilities. This process ensures that masked attention prioritizes CU nodes while excluding irrelevant Rx-AP nodes. Specifically, the masked attention function is defined as
\begin{align}
\label{eq:forward_cu}
&f_{\mathrm{att}}(\mathbf{s}_{\mathrm{Tx}}^{(\tau_1)}(t))=   \nonumber \\  & \underset{K_{\mathrm{u}}}{\mathrm{Top}}\Bigl\{\mathrm{Softmax}\bigl(\underset{\mathcal{N}(t)}{\|}(\mathbf{k}_{\mathcal{N}(t)}^{(\tau_1)})^T\mathbf{W}_{\mathcal{N}(t)}^{\mathrm{att}}\mathbf{q}^{(\tau_1)}(t)+\mathbf{m}_{\mathrm{str}}\bigr)\Bigr\},
\end{align}
where $\mathbf{W}_{\mathcal{N}(t)}^{\mathrm{att}}$ is the attention weight matrix and $\|$ denotes the concatenation operation.} \textcolor{black}{$\mathbf{m}_{\mathrm{str}}$ is the structural mask vector. It employs a masking mechanism (analogous to padded sequence processing) to ignore irrelevant neighbor nodes for each Tx-AP. This ensures that only valid neighbors (specifically, CU nodes) are considered during neighbor aggregation. Concretely, the attention weights corresponding to any Rx-AP neighbors are set to $-\infty$, resulting in zero weight for those neighbors after the softmax operation.} \textcolor{black}{Note that the attention equation integrates the normalized scores of all Tx-AP nodes, and thus is not strictly equivalent to the user association matrix $\bm{\Lambda}$ but rather undergoes a compressed transformation. Specifically, applying the classical Top-$k$ function to the best $K_{\rm u}$ elements of the masked attention equation yields the user association matrix, thereby ensuring a discrete and feasible association structure.}

The query and key vectors are generated through linear projections:
\begin{align}
\mathbf{q}^{(\tau_1)}(t)&=\mathcal{M}^{\mathrm{query}}\left(\mathbf{s}_{\mathrm{Tx}}^{(\tau_1)}(t);\mathbf{W}_{t}^{\mathrm{query}}\right),\\
\mathbf{k}_{\mathcal{N}(t)}^{(\tau_1)}&=\mathcal{M}^{\mathrm{key}}\left(\mathbf{k}_{\mathcal{N}(t)}^{(\tau_1)};\mathbf{W}_{\mathcal{N}(t)}^{\mathrm{key}}\right),
\end{align}
where $\mathcal{M}^{\mathrm{query}}(\cdot)$ and $\mathcal{M}^{\mathrm{key}}(\cdot)$ represent linear transformations with corresponding projection matrices $\mathbf{W}_{t}^{\mathrm{query}}$ and $\mathbf{W}_{\mathcal{N}(t)}^{\mathrm{key}}$, respectively. The node information from neighbors is aggregated similarly to produce value vectors:
\begin{equation}
\underset{\xi\in\mathcal{N}(t)}{f_{\mathrm{mp}}}(\mathbf{s}^{(\tau_1)}(\xi))=\mathcal{M}^{\mathrm{value}}\left(\mathbf{s}^{(\tau_1)}(\xi);\mathbf{W}_{\xi}^{\mathrm{value}}\right).
\end{equation}

Ultimately, the aggregation of structural attention outputs incorporates residual connections to enhance stability and performance:
\begin{align}
f_{\mathrm{agg}}(\mathbf{s}_{\mathrm{Tx}}^{(\tau_1)}(t))&= \mathbf{W}_{t}^{\mathrm{agg}}\mathrm{LeakyReLU}\Bigl\{f_{\mathrm{att}}(\mathbf{s}_{\mathrm{Tx}}^{(\tau_1)}(t)) \\
& \odot\underset{\xi\in\mathcal{N}(t)}{f_{\mathrm{mp}}}(\mathbf{s}^{(\tau_1)}(\xi))\Bigr\}\nonumber+\mathbf{s}_{\mathrm{Tx}}^{(\tau_1)}(t).
\end{align}

However, structural attention alone inadequately determines the AP mode vector $\bm{\kappa}$, as this decision significantly influences both communication sum-rate and target estimation accuracy. Furthermore, using only direct-path interference as prior features limits global radar waveform optimization. To address these limitations, we propose the following scheduling-aware temporal attention module.

\subsubsection{Scheduling-Aware Temporal Attention Module}
Following extraction of the node and edge features from snapshot $\mathcal{G}^{(n)}$, we reconstruct echo signals at each Rx-AP, denoted as $\hat{\mathbf{y}}^{(n)}(r)\in\mathbb{R}^{2N_{\mathrm{s}}LM}$. \textcolor{black}{These signals enhance hidden states for snapshot $(n+1)$:
\begin{align}
\label{enhance1}
\mathbf{s}_{\mathrm{Tx}}^{(\tau^{\mathrm{tot}}_1,n+1)}(t)=\bigl[&\tilde{\mathbf{h}}_{t,1},\ldots,\tilde{\mathbf{h}}_{t,K},\tilde{\mathbf{g}}_{t,1},(\hat{\mathbf{y}}^{(n)}(1))^T,\ldots,\nonumber\\
&\tilde{\mathbf{g}}_{t,R},(\hat{\mathbf{y}}^{(n)}(R))^T\bigr],\\[4pt]
\label{enhance2}
\mathbf{s}_{\mathrm{Rx}}^{(\tau^{\mathrm{tot}}_1,n+1)}(r)=\bigl[&\tilde{\mathbf{h}}_{r,1},\ldots,\tilde{\mathbf{h}}_{r,K},\tilde{\mathbf{g}}_{1,r},(\hat{\mathbf{y}}^{(n)}(r))^T,\ldots,\nonumber\\
&\tilde{\mathbf{g}}_{T,r},(\hat{\mathbf{y}}^{(n)}(r))^T\bigr].
\end{align}}
To maintain dimensional consistency, these hidden states are embedded as $\bar{\mathbf{s}}_{\mathrm{Tx}}^{(\tau^{\mathrm{tot}}_1,n+1)}(t)$ and $\bar{\mathbf{s}}_{\mathrm{Rx}}^{(\tau^{\mathrm{tot}}_1,n+1)}(r)$ between snapshots. \textcolor{black}{By evaluating attention scores between AP node features across adjacent snapshots, the scheduling-aware temporal attention function to optimally select AP modes across snapshots is formulated as:
\begin{align}
\label{eq:forward_ap}
f_{\mathrm{att}}\bigl(\{\bar{\mathbf{s}}^{(\tau^{\mathrm{tot}}_1,n)}\}_{n=1}^{N}\bigr)=\underset{J-R}{\mathrm{Top}}\bigl\{&\underset{\mathcal{N}(j)}{\mathrm{Softmax}}\bigl(\|(\bar{\mathbf{k}}_{\mathcal{N}(j)}^{(\tau^{\mathrm{tot}}_1)})^T\bar{\mathbf{W}}_{\mathcal{N}(j)}^{\mathrm{att}}\bar{\mathbf{q}}^{(\tau^{\mathrm{tot}}_1)}\nonumber\\
&+\mathbf{m}_{\mathrm{tem}}\bigr)\bigr\},
\end{align}
where $\bar{\mathbf{k}}^{(\tau^{\rm tot}_1)}_{\mathcal{N}(j)}$, $\bar{\mathbf{W}}^{\rm att}_{\mathcal{N}(j)}$, and $\bar{\mathbf{q}}^{(\tau^{\rm tot}_1)}$ denote the key vector, attention matrix, and the query vector, respectively.} \textcolor{black}{$\mathbf{m}_{\mathrm{tem}}$ is the temporal mask vector. It functions similarly to $\mathbf{m}_{\mathrm{stru}}$, but in the temporal domain. When a Tx-AP node aggregates features from its neighbors across time (i.e., across consecutive snapshots), $\mathbf{m}_{\rm tem}$ ensures that only features relevant to Rx-AP neighbors are captured, while features from CU neighbors are masked out. The attention weights for CU neighbors are set to $-\infty$, which yields a zero probability for those features after the softmax operation.}
\textcolor{black}{Through similar activation, normalization, and thresholding operations, the framework infers the Tx/Rx configuration for the subsequent snapshot, ensuring temporal consistency and feasibility of the AP mode configuration. The resulting discrete variables $\bm{\Lambda}$ and $\bm{\kappa}$ are determined prior to feature aggregation 
and guide local feature propagation in the dynamic graph construction.}

\textcolor{black}{After processing each snapshot through the structural attention module, the optimized edge features are used to determine the user association matrix $\bm{\Lambda}$, while the node aggregation step refines the precoding matrices $\{\mathbf{W}_{j,i}\}$. To prevent gradient explosion and ensure that the transmit power constraint in \eqref{power_const} is satisfied, each precoding matrix is scaled according to:
\be \mathbf{W}_{j,i} \leftarrow 
\frac{\sqrt{P_{j}}\mathbf{W}_{j,i}}
{\sqrt{\sum_{i=1}^{N_{\rm s}}\|\mathbf{W}_{j,i}\|_F^2}},\ee 
Thus guaranteeing that the power budget at each Tx-AP is always met.}
\vspace{0.1cm}
\subsubsection{3D-Convolution Module}
After scheduling-aware temporal attention-based processing,  the reconstructed echo signals, denoted as $\hat{\mathbf{y}}^{(n)}(r)$, are obtained by concatenating their real and imaginary components along the channel dimension, yielding the tensor $\widehat{\mathbf{Y}}^{(n)} \in \mathbb{R}^{2 \times RM \times N_{\mathrm{s}} \times L}$. Meanwhile, the Rx-APs leverage knowledge of the radar sequence $\mathbf{S}_{\mathrm{r}}$ and the Tx-APs' positions at the $n$-th snapshot to construct another real-valued tensor, denoted by $\widehat{\mathbf{S}}_{\mathrm{r}}^{(n)} \in \mathbb{R}^{2 \times TM \times N_{\mathrm{s}} \times L}$. These two tensors, containing the echo signals and radar sequences respectively, are then fed into the 3D-CNN module.

\textcolor{black}{The deployed 3D-CNN module comprises two components: a feature compression network at each Rx-AP and a centralized feature fusion network at the CPU. At each Rx-AP, a 3D-CNN \cite{CNN} first extracts features from the concatenated input tensor $\{\widehat{\mathbf{Y}}^{(n)}\|\widehat{\mathbf{S}}_{\mathrm{r}}^{(n)}\}\in\mathbb{R}^{2\times JM\times N_{\mathrm{s}}\times L}$. The extracted high-dimensional features are then compressed into a lower-dimensional representation $\mathbf{R}^{(n)} \in \mathbb{R}^{R \times \varrho}$ using multiple linear layers, where $\varrho$ denotes the compressed-feature dimension. Subsequently, the central processing unit (CPU) aggregates these compressed multi-AP features using a multilayer perceptron (MLP) and outputs the estimated positions and velocities  $\{\hat{\mathbf{p}}_{q}^{(n)}, \hat{\mathbf{v}}_{q}^{(n)}\}_{q=1}^Q$. This distributed compression and centralized fusion approach significantly reduces the backhaul signaling overhead required for multi-AP data-level fusion.}

\subsection{Training Strategy}
\label{strategy}
We first generate a training dataset consisting of ground-truth positions $\bar{\mathbf{p}}_{q,\chi}\in\mathbb{R}^{2}$ and velocities $\bar{\mathbf{v}}_{q,\chi}\in\mathbb{R}^{2}$ for each target $q = 1, \dots, Q$, along with their corresponding CSI $\mathbf{H}_{\chi}\in\mathbb{C}^{J\times M \times N_s\times K}$, where $\chi=1,\dots,X$ denotes the dataset sample index and $X$ is the dataset size.

To ensure effective training of the proposed dynamic graph learning framework and to enhance the accuracy of target parameter estimation, we adopt a two-stage training approach. In the first stage, we pre-train the 3D-CNN module, embedded within the receiver signal-processing architecture, using a limited subset of target samples from the dataset. Following this, the parameters from this pre-trained module serve as initialization for the second stage, in which the entire dynamic graph learning framework is trained using the complete dataset.
 
\textcolor{black}{During the initial pre-training stage, the AP mode selection vector $\bm{\kappa}$ is randomly initialized, and the communication objective is temporarily ignored to independently train the 3D-CNN module. At each dataset sample $\chi$, we collect echo signals $\{\widehat{\mathbf{Y}}_{\chi}^{(n)}\}_{n=1}^N$ over $N$ snapshots\footnote{\textcolor{black}{Utilizing only the final snapshot for target estimation during training can cause instability and slow convergence due to feature fluctuation from dynamic graph topology changes. To mitigate this issue, we perform intermediate estimation at every snapshot, providing continuous and stable guidance throughout the training process. At inference, only the final snapshot result is used for estimation.}}, and optimize the mapping function $\mathcal{M}$ to minimize the weighted mean squared error between the estimated target positions/velocities and their corresponding ground-truth values. The pre-training objective is thus formulated as
\vspace{-0.1 cm}
\begin{align}
\mathcal{L}^{(\mathrm{pre})} & = \frac{1}{XNQ}\sum_{\chi=1}^{X} \sum_{n=1}^{N} \sum_{q=1}^{Q}\Bigl\{\omega\|\hat{\mathbf{p}}_{q,\chi}^{(n)} -\bar{\mathbf{p}}_{q,\chi}\|_2^2
\nonumber\\
&\quad+(1-\omega)\|\hat{\mathbf{v}}_{q,\chi}^{(n)} -\bar{\mathbf{v}}_{q,\chi}\|_2^2\Bigr\},
\end{align}
where $\{\hat{\mathbf{p}}_{q,\chi}^{(n)}, \hat{\mathbf{v}}_{q,\chi}^{(n)}\}_{q=1}^{Q} = \mathcal{M}(\widehat{\mathbf{Y}}_{\chi}^{(n)},(\widehat{\mathbf{S}}_{\rm r})_{\chi}^{(n)})$ denotes estimates from the 3D-CNN.}
 
In the subsequent training stage, the pre-trained parameters of the 3D-CNN module are integrated into the dynamic graph learning framework. Here, we jointly optimize the AP mode selection vector $\bm{\kappa}$, user association matrix $\bm{\Lambda}$, precoding matrix $\mathbf{W}_{j,i}$, and the estimate mapping function $\mathcal{M}$ to concurrently maximize the communication sum-rate and minimize estimation errors. By converting constraint \eqref{commun_const} into its Lagrangian equivalent and considering echo signals collected over $N$ snapshots, the joint training objective is formulated as
\begin{align}
\label{loss}
\mathcal{L}^{(\mathrm{train})} &= \frac{1}{X}\sum_{\chi=1}^{X}\Biggl\{\frac{\omega}{NQ}\sum_{n=1}^N\sum_{q=1}^{Q}\|\widehat{\mathbf{p}}_{q, \chi}^{(n)}-\mathbf{p}_{q, \chi}\|_2^2
\nonumber\\
& \hspace{-1.5 cm} +\frac{1-\omega}{NQ}\sum_{n=1}^N\sum_{q=1}^{Q}\|\widehat{\mathbf{v}}_{q, \chi}^{(n)}-\mathbf{v}_{q, \chi}\|_2^2+\rho\left(\gamma-R_{\mathrm{c}}(\mathbf{H}_{\chi})\right)\Biggr\},
\end{align}
where $\rho\in[0,1]$ balances the sensing and communication objectives, $\{\hat{\mathbf{p}}_{q,\chi}^{(n)}, \hat{\mathbf{v}}_{q,\chi}^{(n)}\}_{q=1}^{Q}$ are position and velocity estimates obtained through the 3D-CNN, and $R_{\mathrm{c}}(\mathbf{H}_{\chi})$ denotes the achievable communication sum-rate given the CSI $\mathbf{H}_{\chi}$.  

\textcolor{black}{To enhance training robustness, we employ an adaptive penalty factor $\rho$, progressively increasing its magnitude throughout training. This adaptive strategy guides the optimization towards feasible communication regions, mitigating abrupt constraint violations. In simulation experiments, occasional violations of communication constraints may occur. These instances are addressed by refining initializations and applying a post-correction algorithm~\cite{JPTVT}, ensuring consistent adherence to constraints throughout the inference stage.}

\begin{algorithm}[!t]
\caption{Proposed dynamic graph framework.}
\label{alg:dynamic}
\color{black}
\begin{algorithmic}[1]
\REQUIRE Global CSI, AP positions.
\ENSURE Precoding $\mathbf{W}_{j,i}$, mode vector $\boldsymbol{\kappa}$, association matrix $\boldsymbol{\Lambda}$; estimates $\{\hat{\mathbf{p}}_q,\hat{\mathbf{v}}_q\}_{q=1}^{Q}$.

\FOR{$n=1:\,N$}
    \STATE \textbf{Association pass:}  Each AP performs distributed neighbor aggregation and computes a locally optimal $\boldsymbol{\Lambda}^{(n)}$ and $\boldsymbol{\kappa}^{(n)}$ via the forward propagation in \eqref{eq:forward_cu} and \eqref{eq:forward_ap}.
    \STATE Each AP uploads its association decision to CPU and CPU unifies overall association decision-making.
    \STATE CPU broadcasts global $(\boldsymbol{\Lambda}^{(n)},\boldsymbol{\kappa}^{(n)})$ to APs.
    \STATE \textbf{Precoding pass:} Each Tx-AP performs neighbor aggregation and computes a locally optimal $\mathbf{W}_{j,i}^{(n)}$.
    \STATE Tx-APs transmit using $\mathbf{W}^{(n)}$ with $(\boldsymbol{\Lambda}^{(n)},\boldsymbol{\kappa}^{(n)})$.
    \STATE Rx-APs receive echoes and compress features.
    \STATE Using echoes $\widehat{\mathbf{Y}}^{(n)}$ and  estimator $\mathcal{M}$, obtain $\{\hat{\mathbf{p}}_q^{(n)},\hat{\mathbf{v}}_q^{(n)}\}_{q=1}^{Q}$.
\ENDFOR
\STATE \textbf{Return} $\mathbf{W}_{j,i}, \boldsymbol{\kappa}, \boldsymbol{\Lambda}$ and the estimates $\{ \hat{\mathbf{p}}_q , \hat{\mathbf{v}}_q \}_{q=1}^{Q}$.
\end{algorithmic}
\end{algorithm}

\color{black}
\subsection{Backhaul Signaling Overhead Analysis}
In this subsection, we quantitatively evaluate the backhaul signaling overhead associated with the proposed dynamic graph learning framework. Specifically, overhead is measured in terms of the number of double-precision floating-point values exchanged between the CPU and distributed APs. Typically, backhaul signaling comprises global CSI distribution, AP-user association data, precoding matrices, inter-AP coordination messages, and compressed echo-signal features. The full operational procedure of the proposed framework is summarized in Algorithm~\ref{alg:dynamic}.

Initially, at each snapshot, the CPU distributes the global CSI to the APs, requiring approximately $J M N_{\rm s} K$ double-precision values. Defining $D_{\mathrm{AP}}$ and $D_{\mathrm{CU}}$ as dimensions of the aggregation weight matrices for AP and CU nodes, respectively. During attention-based association inference, each AP computes attention scores by processing $\tau^{\rm tot}_1 D_{\rm AP} N_{\rm s} M J^2 + \tau^{\rm tot}_1 D_{\rm CU} N_{\rm s} M K$ double-precision values. Subsequently, each AP uploads its local association decision, amounting to around $J^2 + J^2 K$ double-precision values. The CPU then broadcasts the consolidated global association decisions back to the APs, requiring an additional $J + T K$ double-precision values.

Following the establishment of graph edge weights, each Tx-AP performs local feature aggregation for precoding optimization. This step involves approximately $\tau^{\rm tot}_1 T(D_{\rm Rx} N_{\rm s} M R + D_{\rm CU} N_{\rm s} M K_{\rm u})$ double-precision values. The optimized precoding matrices are computed and stored locally at each Tx-AP.

Finally, Rx-APs upload compressed echo-signal features to the CPU, significantly reducing the backhaul overhead. Specifically, transmitting these compressed features requires only $R \varrho$ double-precision values, substantially lower than the original data size of $2 R M N_{\rm s} L$.

\color{black}

\section{Proposed Mirror-GAT Framework}
Despite employing feature compression, the dynamic graph learning framework proposed in the previous section inherently involves substantial overhead due to frequent AP-CPU exchanges and complex interactions across snapshots. To address these practical constraints, we introduce the mirror-GAT framework, explicitly designed to reduce both computational complexity and backhaul signaling overhead. In contrast to the dynamic graph approach, mirror-GAT adopts a single pair of shared heterogeneous GAT modules, significantly reducing the number of trainable parameters and facilitating scalable deployment in large-scale cell-free ISAC networks. To effectively leverage this structure, we decompose the original joint optimization into two interconnected yet tractable subproblems: (\textit{i}) precoding optimization and (\textit{ii}) AP mode selection and CU association, solved iteratively by adjacency-shared GAT modules. The detailed methodology is presented in the following subsections.

\subsection{Problem Transformation}
\label{task}

To facilitate structured and independent optimization within the mirror-GAT framework, we first integrate the pre-trained 3D-CNN module (described in Section~\ref{strategy}) as a fixed signal-level fusion mapping. By treating the 3D-CNN module as a fixed estimator, we exclude it from the optimization variables, thereby significantly reducing the complexity and dimensionality of the optimization problem. With this fixed estimator in place, we focus explicitly on jointly optimizing AP mode selection, user association, and transmit precoding matrices.

For improved tractability, we decompose the original problem into two interdependent, hierarchical subproblems. Specifically, given a fixed AP mode selection vector $\bm{\kappa}^\star$ and user association matrix $\bm{\Lambda}^\star$, the lower-level subproblem concentrates exclusively on optimizing the precoding matrices. Formally, this precoding optimization subproblem is expressed as:
\begin{subequations}
\label{first_stage}
\begin{align}
&\underset{\mathbf{W}_{j,i}}{\min}\quad
\omega\mathrm{RMSE}_{\text{p}}+(1-\omega)\mathrm{RMSE}_{\text{v}}\\
&\mathrm{s.t.}\quad R_\mathrm{c} \geq\gamma,\\
&\quad\quad~\sum\limits_{i=1}^{N_{\mathrm{s}}}\|\mathbf{W}_{j,i}\|_{F}^{2}\leq P_j, ~\forall j\in\mathcal{T}.
\end{align}
\end{subequations}
Using the optimized precoding matrices derived from the solution of the lower-level problem \eqref{first_stage}, we then formulate the upper-level subproblem, aimed at optimizing the AP mode selection and user association decisions:
\begin{subequations}
\label{second_stage}
\begin{align}
&\underset{\bm{\Lambda},\bm{\kappa}}{\min}\quad \omega\mathrm{RMSE}_{\text{p}}+(1-\omega)\mathrm{RMSE}_{\text{v}}\\
&\mathrm{s.t.}\quad R_\mathrm{c} \geq \gamma.
\end{align}
\end{subequations}
This bi-level reformulation decouples the original joint optimization task into two more tractable subproblems, allowing each GAT module to efficiently solve a specific subtask with reduced computational complexity. Through iterative alternation between these two optimization steps, the mirror-GAT framework systematically converges to locally optimal solutions for the precoding matrices $\mathbf{W}_{j,i}^\star$, AP mode selection vector $\bm{\kappa}^\star$, and user association matrix $\bm{\Lambda}^\star$. The detailed implementation and iterative inference mechanism are described next.

\subsection{Graph Definition}
Following problem transformation, we establish two heterogeneous mirror graphs, denoted by $\mathcal{G}_{1}=(\mathcal{V}_1,\mathcal{E})$ and $\mathcal{G}_2=(\mathcal{V}_2,\mathcal{E})$, to iteratively optimize the precoding matrix  $\mathbf{W}_{j,i}$, the AP mode selection vector $\bm{\kappa}$, and the user association matrix $\bm{\Lambda}$.
The two mirror graphs share the same adjacency structure but fundamentally differ in message-passing directions and optimization objectives. Specifically, the association-oriented graph $\mathcal{G}_2$ employs node-to-edge message passing to optimize AP mode selection and user association, while the precoding-oriented graph $\mathcal{G}_1$ utilizes edge-to-node message passing for optimizing precoding matrices.

\subsubsection{Precoding-Oriented Graph}
The node set $\mathcal{V}_1$ of the precoding-oriented graph comprises three distinct node types: Tx-AP, Rx-AP, and CU, formally expressed as $\mathcal{V}_1=\{\mathcal{V}_{\mathrm{Tx}},\mathcal{V}_{\mathrm{Rx}},\mathcal{V}_{\mathrm{CU}}\}$. Edges in this graph represent communication and sensing interactions, with edge features derived from the reshaped channel features defined in \eqref{eq:htk} and \eqref{eq:Gtr}.

In the precoding optimization task, the node features of each Tx-AP are iteratively propagated and updated through $\tau_1^{\mathrm{tot}}$ hidden layers, using the following message-passing framework:
\begin{equation}
\label{forward}
\mathbf{s}^{(\tau_{2}+1)}_{\mathrm{Tx}}(t)\!=\!h_{\mathrm{agg}}\bigl(h_{\mathrm{att}}(\mathbf{s}^{(\tau_{2})}_{\mathrm{Tx}}(t))\odot h_{\mathrm{mp}}(\{\mathbf{s}^{(\tau_{2})}(\xi)\}_{\xi\in\mathcal{N}(t)})\bigr),
\end{equation}
where $h_{\mathrm{agg}}(\cdot)$, $h_{\mathrm{att}}(\cdot)$, and $h_{\mathrm{mp}}(\cdot)$ denote aggregation, attention scoring, and message passing functions, respectively.

In the precoding design task, we assume that the framework has prior knowledge of the AP modes and user association strategies. Unlike the proposed dynamic graph learning framework (which extracts masked attention scores from edge information), the precoding-oriented graph utilizes a weighted attention scoring criterion for enhanced feature aggregation. Specifically, the attention mechanism is defined as:
\begin{align}
h_{\textrm{att}}(\mathbf{s}^{(\tau_2)}_{\textrm{Tx}}(t))&= \alpha \textrm{Softmax}\big(\underset{\mathcal{N}(t)}{\|}(\mathbf{k}^{(\tau_2)}_{\mathcal{N}(t)})^T\mathbf{W}^{\rm att}_{\mathcal{N}(t)}\mathbf{q}^{(\tau_2)}(t)\big)\nonumber \\
&+(1-\alpha)\{(\mathbf{1}-\bm{\kappa})^T\|\bm{\Lambda}(t,:)\},
\end{align}
where $\alpha$ is the weighting factor for the attention score, $\mathbf{W}^{\rm att}_{\mathcal{N}_(t)}$, $\mathbf{k}_{\mathcal{N}(t)}^{(\tau_2)}$, and $\mathbf{q}^{(\tau_2)}$ denote the attention weight matrix, query vector, and key vector, respectively.

Subsequently, the node information from neighboring nodes is aggregated into value vectors using a message-passing function:
\be
\underset{\xi\in\mathcal{N}(t)}{h_{\textrm{mp}}}(\mathbf{s}^{(\tau_2)}(\xi))=\mathcal{M}^{\rm value}(\mathbf{s}^{(\tau_2)}(\xi);\mathbf{W}^{\rm value}_{\xi}).
\ee

Finally, the aggregation stage incorporates residual connections to ensure stability, mitigate gradient vanishing, and improve convergence during training:
\begin{align}
h_{\textrm{agg}}(\mathbf{s}^{(\tau_2)}_{\textrm{Tx}}(t))&=\mathbf{W}^{\rm agg}_t\textrm{LeakyReLU}\big\{h_{\textrm{att}}(\mathbf{s}^{(\tau_2)}_{\textrm{Tx}}(t)) \nonumber
\\
&\odot\underset{\xi\in\mathcal{N}(t)}{h_{\textrm{mp}}}(\mathbf{s}^{(\tau_2)}(\xi))\big\}\!+\!\mathbf{s}^{(\tau_2)}_{\textrm{Tx}}(t).
\end{align}
By combining weighted attention scoring and residual connections, each Tx-AP node effectively aggregates neighboring features to determine locally optimal precoding matrices $\mathbf{W}_{j,i}^\star$. Thus, the precoding-oriented graph explicitly solves the lower-level optimization subproblem \eqref{first_stage}.

\subsubsection{Association-Oriented Graph}
Due to inherent uncertainties associated with AP mode selection, the association-oriented graph is designed with a simplified node structure comprising only two types of nodes: AP nodes and CU nodes. Within this structure, dynamic connections among AP nodes are reconfigured adaptively, and weighted communication links between AP and CU nodes are dynamically established based on their channel characteristics and mutual interference conditions. Such a dynamic association-oriented approach allows the simultaneous optimization of AP mode selection and user associations. Formally, the edge set $\mathcal{E}$ is partitioned into two distinct subsets: communication edges ($\mathcal{E}_{\mathrm{com}}$), connecting AP nodes directly to CU nodes, and sensing edges ($\mathcal{E}_{\mathrm{sen}}$), representing inter-AP sensing interactions. Hence, the complete edge set is expressed as $\mathcal{E} = \{\mathcal{E}_{\mathrm{com}}, \mathcal{E}_{\mathrm{sen}}\}$.

Given prior knowledge of multi-user CSI at each AP and corresponding locally optimized precoding matrices $\mathbf{W}_{j,i}$ obtained from the precoding-oriented graph, we initialize the hidden states of edges in terms of normalized precoded channel gains as follows:
\begin{align}
\mathbf{s}^{(0)}_{\mathrm{com}}(j,k)&=\sum_{i=1}^{N_{\mathrm{s}}}\mathbf{h}_{j,i,k}^H\mathbf{w}_{\text{c},j,i,k},\\
\mathbf{s}^{(0)}_{\mathrm{sen}}(j,\ell)&=\frac{1}{M^2}\sum_{n=1}^M\sum_{i=1}^{N_{\mathrm{s}}}\sum_{m=1}^{M}\frac{1}{[\mathbf{G}_{j,\ell}\mathbf{w}_{\text{r},j,i,m}]_n},
\end{align}
where $\ell\in\mathcal{J}\setminus\{j\}$.

Subsequently, each edge hidden state is processed through a dedicated heterogeneous GAT consisting of $\tau_{2}^{\mathrm{tot}}$ hidden layers. The forward propagation updates within the association-oriented graph are formulated as:
\begin{align}
\label{fp_2}
\mathbf{s}^{(\tau_{2}+1)}_{\mathrm{com}}(j,k)&=g_{\mathrm{agg}}\Bigl(g_{\mathrm{att}}\big(\mathbf{s}^{(\tau_{2})}_{\mathrm{com}}(j,k)\big)\odot \underset{\epsilon \in\mathcal{N}(j,k)}{g_{\mathrm{mp}}}\big(\mathbf{s}^{(\tau_{2})}(\epsilon)\big)\Bigr),\\
\mathbf{s}^{(\tau_{2}+1)}_{\mathrm{sen}}(j,\ell)&=g_{\mathrm{agg}}\Bigl(g_{\mathrm{att}}\big(\mathbf{s}^{(\tau_{2})}_{\mathrm{sen}}(j,\ell)\big)\odot \underset{\epsilon \in\mathcal{N}(j,\ell)}{g_{\mathrm{mp}}}\big(\mathbf{s}^{(\tau_{2})}(\epsilon)\big)\Bigr),
\end{align}
where $g_{\mathrm{agg}}(\cdot)$, $g_{\mathrm{att}}(\cdot)$, and $g_{\mathrm{mp}}(\cdot)$ respectively represent feature aggregation, edge-based attention scoring, and edge-to-edge message passing functions. Here, $\mathcal{N}(j,\ell)$ denotes the set of neighboring edges involved in message passing for the edge $(j,\ell)$.

To effectively guide AP-user association and AP mode selection, the attention scoring functions explicitly incorporate global node-weight information:
\begin{align}
g_{\mathrm{att}}\big(\mathbf{s}^{(\tau_2)}_{\mathrm{com}}(j,k)\big)&=\underset{\mathcal{N}(j,k)}{\mathrm{Softmax}}\bigl(\|(\mathbf{k}^{(\tau_2)}_{\mathcal{N}(j,k)})^T\mathbf{W}^{\mathrm{att}}_{\mathcal{N}(j,k)}\mathbf{q}^{(\tau_2)}(j,k)\nonumber\\
&\qquad\qquad\quad~+\mathbf{u}_{j,k}^T\mathbf{s}_{\mathrm{com}}^{(\tau_2)}\bigr),\\
g_{\mathrm{att}}\big(\mathbf{s}^{(\tau_2)}_{\mathrm{sen}}(j,\ell)\big)&=\underset{\mathcal{N}(j,\ell)}{\mathrm{Softmax}}\bigl(\|(\mathbf{k}^{(\tau_2)}_{\mathcal{N}(j,\ell)})^T\mathbf{W}^{\mathrm{att}}_{\mathcal{N}(j,\ell)}\mathbf{q}^{(\tau_2)}(j,\ell)\nonumber\\
&\qquad\qquad\quad~+\mathbf{u}_{j,\ell}^T\mathbf{s}_{\mathrm{sen}}^{(\tau_2)}\bigr),
\end{align}
where $\mathbf{W}^{\mathrm{att}}_{\mathcal{N}}$, $\mathbf{k}^{(\tau_2)}_{\mathcal{N}}$, and $\mathbf{q}^{(\tau_2)}$ respectively denote the attention weight matrices, key vectors, and query vectors utilized for edge-based attention scoring. The vector $\mathbf{u}$ contains trainable parameters that enhance edge features during the attention evaluation phase.
Next, the attention-weighted information from neighboring edges is aggregated via message passing to yield value vectors:
\begin{align}
\underset{\epsilon\in\mathcal{N}}{g_{\mathrm{mp}}}\big(\mathbf{s}^{(\tau_2)}(\epsilon)\big)=\mathcal{M}^{\mathrm{value}}\big(\mathbf{s}^{(\tau_2)}(\epsilon);\mathbf{W}^{\mathrm{value}}_{\epsilon}\big).
\end{align}
Finally, the local edge-feature aggregation stage employs residual connections to stabilize training, improve network performance, and refine feature representations. The aggregated edge states are updated as:
\begin{align}
&g_{\mathrm{agg}}\big(\mathbf{s}^{(\tau_2)}_{\mathrm{com}}(j,k)\big)=\mathbf{s}^{(\tau_2)}_{\mathrm{com}}(j,k)\nonumber\\
&+\mathbf{W}^{\mathrm{agg}}_{j,k}\mathrm{Softmax}\Bigl\{g_{\mathrm{att}}\big(\mathbf{s}^{(\tau_2)}_{\mathrm{com}}(j,k)\big)\odot\underset{\epsilon\in\mathcal{N}(j,k)}{g_{\mathrm{mp}}}\big(\mathbf{s}^{(\tau_2)}(\epsilon)\big)\Bigr\},\\
&g_{\mathrm{agg}}\big(\mathbf{s}^{(\tau_2)}_{\mathrm{sen}}(j,\ell)\big)=\mathbf{s}^{(\tau_2)}_{\mathrm{sen}}(j,\ell)\nonumber\\
&+\mathbf{W}^{\mathrm{agg}}_{j,\ell}\mathrm{Softmax}\Bigl\{g_{\mathrm{att}}\big(\mathbf{s}^{(\tau_2)}_{\mathrm{sen}}(j,\ell)\big)\odot\underset{\epsilon\in\mathcal{N}(j,\ell)}{g_{\mathrm{mp}}}\big(\mathbf{s}^{(\tau_2)}(\epsilon)\big)\Bigr\}.
\end{align}

By executing this structured edge-to-edge information integration, the association-oriented graph effectively computes globally optimized edge weights. Consequently, these edge features enable the determination of the locally optimal AP mode selection vector $\bm{\kappa}^\star$ and user association matrix $\bm{\Lambda}^\star$, solving the upper-level optimization subproblem~\eqref{second_stage}.

\subsection{Mirror-based Iteration}
\begin{figure}[!t]
\centering
\includegraphics[width=3.4 in]{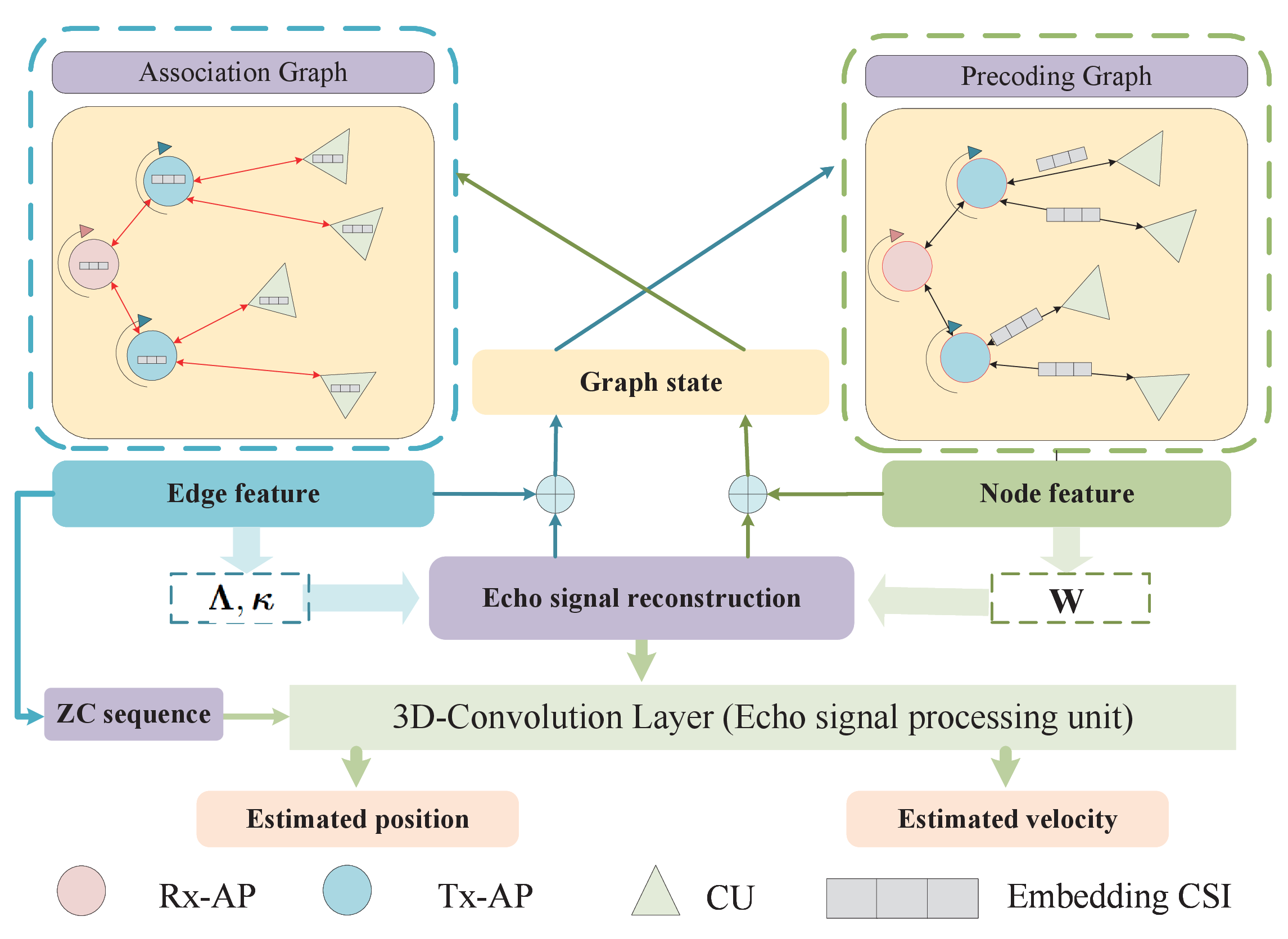}
\caption{Proposed mirror-GAT framework.}\vspace{-0.2 cm}
\label{mirror}
\end{figure}

While the precoding-oriented and association-oriented graphs individually yield locally optimal solutions, executing each graph only once may lead to convergence toward suboptimal local minima due to the inherently coupled structure of the optimization problem. Additionally, the mirror-GAT framework operates without prior knowledge of target parameters, rendering single-pass inference insufficient for achieving high-precision estimation. To address these limitations, we propose a mirror-based iterative refinement strategy that cyclically enhances the solutions by incorporating reconstructed echo signals as feedback, as illustrated in Fig.~\ref{mirror}.

Let $N_{\mathrm{m}}$ denote the total number of mirror iterations. During the $n_{\mathrm{m}}$-th mirror iteration, both the precoding-oriented graph $\mathcal{G}_1^{(n_{\mathrm{m}})}$ and the association-oriented graph $\mathcal{G}_2^{(n_{\mathrm{m}})}$ perform forward inference independently. Consequently, the mirror-GAT framework outputs the locally optimal precoding matrices $\mathbf{W}_{j,i}^{(n_{\mathrm{m}})}$, AP mode selection vector $\bm{\kappa}^{(n_{\mathrm{m}})}$, and user association matrix $\bm{\Lambda}^{(n_{\mathrm{m}})}$ at each iteration.

Following extraction of node and edge features and determination of local solutions, the echo signal tensor $\widehat{\mathbf{Y}}^{(n_{\mathrm{m}})}\in\mathbb{C}^{R\times M\times N_{\mathrm{s}}\times L}$ is reconstructed at each Rx-AP. This tensor is then reshaped into feature vectors $\hat{\mathbf{y}}^{(n_{\mathrm{m}})}(r)\in\mathbb{R}^{2N_{\mathrm{s}}LM}$ for each Rx-AP node $r\in\mathcal{R}^{(n_{\mathrm{m}})}$, where $\mathcal{R}^{(n_{\mathrm{m}})}$ denotes the set of Rx-AP nodes in the $n_{\mathrm{m}}$-th iteration. For all $n_{\mathrm{m}}>1$, these echo features are utilized to update the initial hidden states for the subsequent iteration as follows:
\begin{align}
\label{update_1}
\mathbf{s}^{(n_{\mathrm{m}}+1, 0)}_{\mathrm{Tx}}(t)=&\bigl[\tilde{\mathbf{h}}_{t,1},\ldots,\tilde{\mathbf{h}}_{t,K_{\rm u}},\tilde{\mathbf{g}}_{t,1},(\hat{\mathbf{y}}^{(n_{\mathrm{m}})}(r))^T,\ldots,\nonumber\\
& \hspace{1 cm} \tilde{\mathbf{g}}_{t,R},(\hat{\mathbf{y}}^{(n_{\mathrm{m}})}(R))^T\bigr],
\end{align}
\begin{align}
\label{update_2}
\mathbf{s}^{(n_{\mathrm{m}}+1, 0)}_{\mathrm{Rx}}(r)=&\bigl[\tilde{\mathbf{g}}_{1,r},(\hat{\mathbf{y}}^{(n_{\mathrm{m}})}(r))^T,\ldots,  \tilde{\mathbf{g}}_{T,r},(\hat{\mathbf{y}}^{(n_{\mathrm{m}})}(r))^T\bigr],
\end{align}
\begin{align}
\label{update_3}
\mathbf{s}^{(n_{\mathrm{m}}+1, 0)}_{\mathrm{sen}}(j,\ell )\!=\!\frac{1}{M^2}\!\sum_{n,i,m}\!\frac{\displaystyle\sum_{l}\widehat{\mathbf{Y}}^{(n_{\mathrm{m}})}(\ell,m,i,l)}{[\mathbf{G}_{j,\ell}\mathbf{w}_{\text{r},j,i,m}]_n}\!+\!\mathbf{s}^{(n_{\mathrm{m}}, 0)}_{\mathrm{sen}}(j,\ell ).
\end{align}
Following principles analogous to those presented in \eqref{enhance1} and \eqref{enhance2}, an embedding operation is applied between successive iterations to maintain dimensional consistency and ensure seamless feature propagation across iterations.

\subsection{Training Strategy}
\label{TS:2}
The training and testing phases of the mirror-GAT framework employ the same ground-truth datasets and pre-trained 3D-CNN modules as described in Section~\ref{strategy}. To accelerate convergence of the iterative process, we initialize the precoding-oriented graph $\mathcal{G}_1^{(0)}$ using a hierarchical proximity-based strategy.

In the absence of target position information, the initial AP mode selection vector $\bm{\kappa}^{(0)}$ is formed by selecting the $R$ AP nodes with the largest aggregate distances from all other APs:
\begin{align}
    \bm{\kappa}^{(0)}=\underset{R}{\mathrm{Top}}\left\{\sum_{s=1,s\neq j}^{J}d_{:,s}\right\}.
\end{align}
After determining $\bm{\kappa}^{(0)}$, each AP node selects the  $K_{\mathrm{u}}$ user nodes with the strongest channel gains as its initial association vector:
\begin{align}    \bm{\Lambda}_{j,:}^{(0)}=\underset{K_{\mathrm{u}}}{\mathrm{Top}}\left\{\sum_{i=1}^{N_{\mathrm{s}}}\left\|\mathbf{h}_{j,i,:}\right\|^2\right\}.
\end{align}
Using these initializations, the mirror-GAT framework iteratively computes the optimal precoding matrices $\mathbf{W}_{j,i}$, along with the optimal AP mode selection vector $\bm{\kappa}$ and user association matrix $\bm{\Lambda}$.

\begin{algorithm}[!t]
\caption{Proposed mirror-GAT framework.}
\label{alg:mirror_GAT}
\color{black}
\begin{algorithmic}[1]
\REQUIRE Global CSI; AP positions.
\ENSURE Precoding $\mathbf{W}_{j,i}$, mode vector $\boldsymbol{\kappa}$, association matrix $\boldsymbol{\Lambda}$; estimates $\{\hat{\mathbf{p}}_q,\hat{\mathbf{v}}_q\}_{q=1}^{Q}$.
\STATE \textbf{Initialization:} CPU generates $\boldsymbol{\Lambda}^{(0)}$ and $\boldsymbol{\kappa}^{(0)}$.
\FOR{$n_{\rm m}=1:\,N_{\rm m}$}
    \STATE CPU broadcasts current $(\boldsymbol{\Lambda}^{(n_{\rm m}-1)},\boldsymbol{\kappa}^{(n_{\rm m}-1)})$ and CSI to Tx-APs.
    \STATE \textbf{Precoding pass:} Each Tx-AP performs distributed neighbor aggregation and computes a locally optimal $\mathbf{W}_{j,i}^{(n_{\rm m})}$ via forward propagation in \eqref{forward}.
    \STATE Tx-APs transmit with $\mathbf{W}^{(n_{\rm m})}$ under $(\boldsymbol{\Lambda}^{(n_{\rm m}-1)},\boldsymbol{\kappa}^{(n_{\rm m}-1)})$ and upload designed precoders to CPU.
    \STATE Rx-APs receive echoes, form $\widehat{\mathbf{Y}}^{(n_{\rm m})}$, and compress features for association-oriented pass.
    \STATE \textbf{Association pass:} CPU runs  association-oriented GAT forward in \eqref{fp_2} using global CSI and compressed echoes, yielding $(\boldsymbol{\kappa}^{(n_{\rm m})},\boldsymbol{\Lambda}^{(n_{\rm m})})$.
\ENDFOR
\STATE Using final echo $\widehat{\mathbf{Y}}^{(N_{\rm m})}$ and  estimator $\mathcal{M}$, obtain $\{\hat{\mathbf{p}}_q,\hat{\mathbf{v}}_q\}_{q=1}^{Q}$.
\STATE \textbf{Return} $\mathbf{W}_{j,i}, \boldsymbol{\kappa}, \boldsymbol{\Lambda}$ and $\{\hat{\mathbf{p}}_q,\hat{\mathbf{v}}_q\}_{q=1}^{Q}$.
\end{algorithmic}
\end{algorithm}

\color{black}
\subsection{Backhaul Signaling Overhead Reduction Analysis}
We now quantitatively assess the backhaul signaling overhead of the proposed mirror-GAT framework, highlighting its benefits relative to the dynamic graph learning approach. The mirror-GAT explicitly separates the association optimization (performed centrally by the CPU) from precoding optimization (performed locally by distributed Tx-APs). The detailed operational procedure is summarized in Algorithm~\ref{alg:mirror_GAT}. During each mirror iteration, the CPU transmits approximately $J + T K$ double-precision values for global association data and $T M N_{\mathrm{s}} K_{\mathrm{u}}$ values for local CSI to Tx-APs. Each Tx-AP computes and uploads precoding matrices, requiring roughly $T M N_{\mathrm{s}}(K_{\mathrm{u}} + M)$ double-precision values.

Let $D_{\mathrm{Rx}}$ and $D_{\mathrm{CU}}$ denote the dimensions of the aggregation weight matrices for Rx-AP and CU nodes, respectively. Each Tx-AP consumes about $\tau^{\mathrm{tot}}_2 T(D_{\mathrm{Rx}}N_{\mathrm{s}}M R + D_{\mathrm{CU}} N_{\mathrm{s}}M K_{\mathrm{u}})$ double-precision values during message passing. The Rx-APs upload compressed echo-signal features to the CPU, significantly reducing overhead, requiring only $R \varrho$ double-precision values per iteration.

Overall, the mirror-GAT significantly reduces backhaul signaling overhead by clearly separating tasks, limiting communications to a single exchange per iteration, minimizing precoding data transmissions, and avoiding bi-directional synchronization. Thus, it provides superior scalability and practical feasibility compared to the dynamic graph learning framework in large-scale cooperative cell-free ISAC deployments.

\color{black}

\section{Simulation Results}

\textcolor{black}{Unless otherwise specified, the simulation parameters and configurations are as follows. The considered cooperative cell-free ISAC system consists of $14$ Tx-APs and $2$ Rx-APs, each equipped with $8$ antennas. Each AP has a transmit power budget of $50$ dBm. The carrier frequency is $24$ GHz, and the system bandwidth is $120$ MHz with $32$ subcarriers and $16$ OFDM symbols. The Tx-APs simultaneously serve $16$ single-antenna CUs and sense $3$ targets with velocities uniformly distributed from $-30$ to $30$ m/s. Each Tx-AP associates with at most $K_{\mathrm{u}}=2$ CUs. Communication and sensing noise powers are both set at $-80$ dBm, and the path-loss exponent is uniformly distributed between $2.1$ and $3.8$. The GNN comprises $10$ hidden layers. During training, the system uses $10$ temporal snapshots and $10$ mirror iterations, with compressed‑feature dimension $\varrho=25$. The training dataset includes $12,000$ samples, and the learning rate is $0.0001$.} 

\vspace{-0.2 cm}
\subsection{Illustration of AP Selection and CU Association}
\begin{figure}[!t]
\centering
\includegraphics[width=3.2 in]{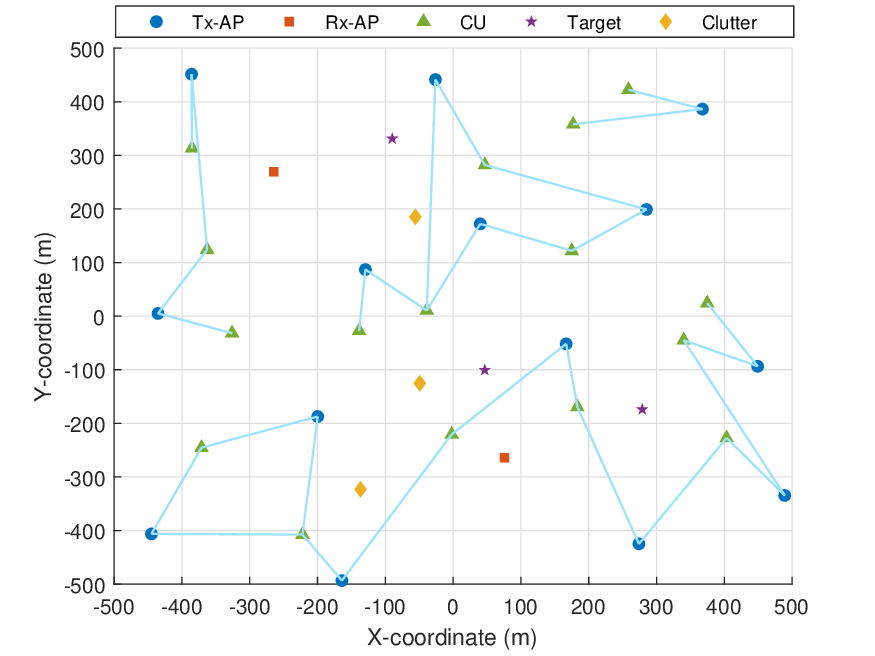}
\caption{Illustration of AP mode selection and user association.}
\label{scatter}
\vspace{-0.3 cm}
\end{figure}

Fig.~\ref{scatter} illustrates the AP mode selection and CU association results obtained by the proposed dynamic graph learning framework. In the figure, network entities are marked as follows: Tx-APs (blue circles), Rx-APs (red squares), CUs (green triangles), sensing targets (purple pentagrams), and clutter sources (yellow diamonds). The established communication links between Tx-APs and their associated CUs are depicted by blue connecting lines. It is observed that the proposed method adaptively selects AP operating modes and forms efficient AP-CU associations. Notably, APs positioned close to sensing targets yet sufficiently distant from clutter sources are preferentially configured as Rx-APs. This spatially adaptive selection significantly improves the quality of received echo signals, mitigates clutter interference, and consequently enhances the accuracy of multi-target position and velocity estimation.

\vspace{-0.2 cm}

\subsection{Position and Velocity Estimation Performance}
\label{Performance}
We now evaluate the position and velocity estimation performance of the proposed graph-learning frameworks. To highlight the advantages of our dynamic graph learning (\textbf{Prop. dynamic graph}) and mirror-GAT (\textbf{Prop. mirror-GAT}) methods, we compare them with the following baseline schemes:
\begin{itemize}
\item \textbf{Heuristic+GAT}: Uses the hierarchical proximity principle (see Sec.~\ref{TS:2}) for heuristic AP mode selection and user association within the mirror-GAT structure.
\item \textbf{Random+GAT}: Randomly selects AP modes and user associations within the mirror-GAT framework.
\item \textcolor{black}{\textbf{MMSE+CNN} \cite{WZHICASSP}: Employs an MMSE precoder combined with heuristic association, using a centralized 3D-CNN model for signal-level fusion and parameter estimation without prior target information.}
\item \textcolor{black}{\textbf{ZF+FFT} \cite{XZCTSP2024}: Adopts a zero-forcing precoder and heuristic association, performing least-squares fusion via the classical 3D-FFT method for position and velocity estimation without prior target knowledge.}
\item \textcolor{black}{\textbf{Two-stage} \cite{Two_stage}: Initially utilizes wide-beam transmissions to compute range-angle maps at each Rx-AP, followed by cooperative maximum likelihood-based refinement of positioning and velocity estimates.}
\end{itemize}

\begin{figure}[!t]
\centering
\subfloat[Position RMSE]{
		\includegraphics[width=1.6 in,scale=0.5]{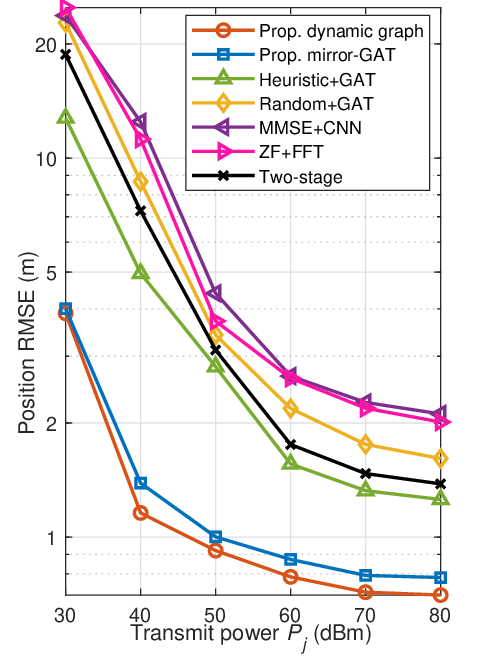}}
\subfloat[Velocity RMSE]{
		\includegraphics[width=1.6 in,scale=0.5]{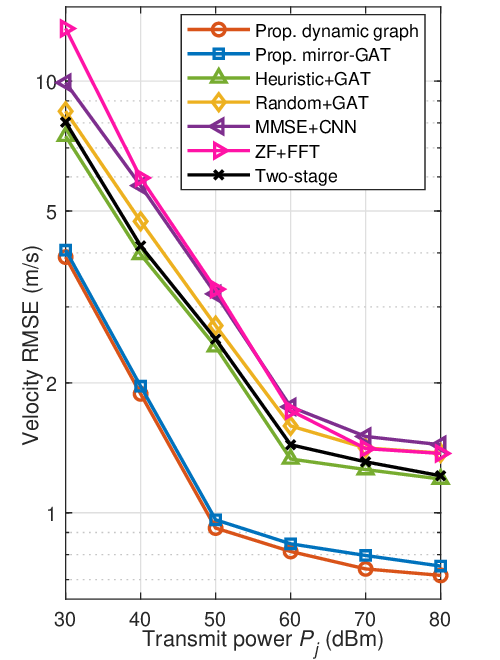}}
\caption{Sensing performance versus transmit power $P_j$.}\vspace{-0.2 cm}
\label{Sim:RMSE_P}
\end{figure}

Fig.~\ref{Sim:RMSE_P}(a) and Fig.~\ref{Sim:RMSE_P}(b) illustrate the RMSE performance of target position and velocity estimations, respectively, as functions of per-AP transmit power $P_j$. As expected, increasing $P_j$ consistently improves estimation accuracy across all methods. The proposed dynamic graph and mirror-GAT frameworks significantly outperform the heuristic and random baselines, achieving position and velocity RMSEs below $1$m and $1$m/s, respectively, at moderate transmit power (e.g., around 50-60 dBm in Fig. 6). \textcolor{black}{Meanwhile, our proposed methods consistently surpass the conventional MMSE+CNN, ZF+FFT, and two-stage approaches, clearly demonstrating the advantages of our end-to-end robust design in cooperative network optimization, adaptive precoding, and target parameter estimation without prior knowledge.} Furthermore, the dynamic graph framework achieves slightly better performance than the mirror-GAT framework, but at the cost of higher computational complexity and signaling overhead (to be analyzed later).

\begin{figure}[!t]
\centering
\subfloat[Position RMSE]{
		\includegraphics[width=1.7 in,scale=0.5]{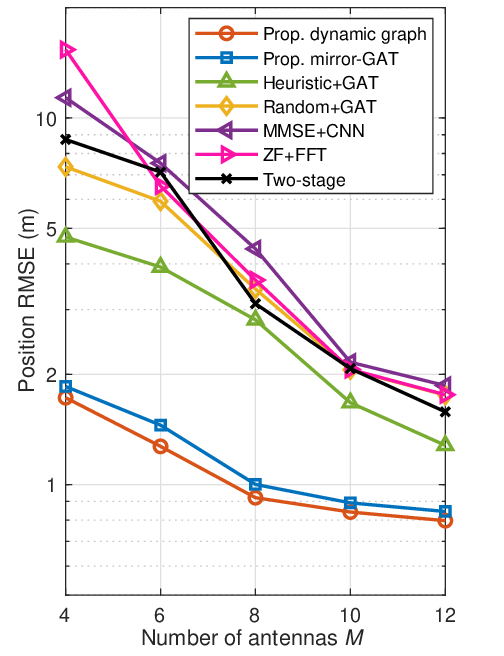}}
\subfloat[Velocity RMSE]{
		\includegraphics[width=1.7 in,scale=0.5]{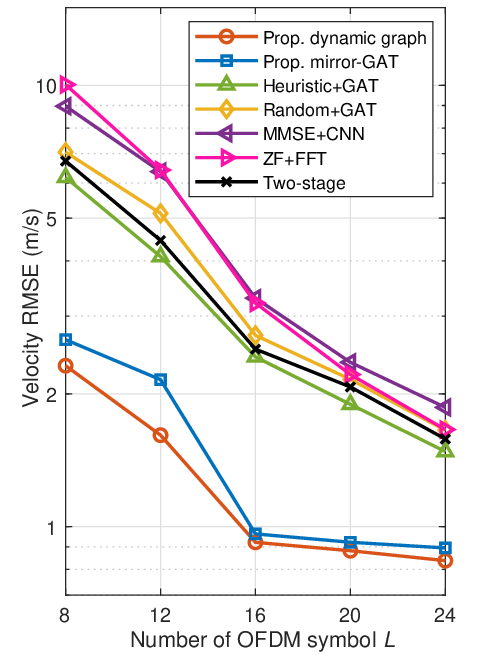}}
\caption{Sensing performance under different parameters.}\vspace{0 cm}
\label{Sim:RMSE_ML}
\vspace{-0.3cm}
\end{figure}

Fig.~\ref{Sim:RMSE_ML} evaluates the impact of the number of transmit antennas $M$ and OFDM symbols $L$ on estimation performance. Fig.~\ref{Sim:RMSE_ML}(a) demonstrates that increasing $M$ significantly improves positioning accuracy for both the dynamic graph and mirror-GAT frameworks, primarily due to higher spatial resolution and array gain. Similarly, Fig.~\ref{Sim:RMSE_ML}(b) shows that increasing $L$ enhances velocity estimation accuracy, attributed to the extended coherent processing interval and improved Doppler resolution. Notably, the proposed frameworks exhibit strong generalization capability, achieving excellent estimation performance even under testing configurations that differ from those used during training. Moreover, across all tested configurations, both proposed methods consistently outperform the heuristic and random baselines, confirming the effectiveness of the joint graph-learning-based optimization and estimation approach.

\begin{figure}[!t]
\centering
\subfloat[Position RMSE]{
		\includegraphics[width=1.75 in,scale=0.5]{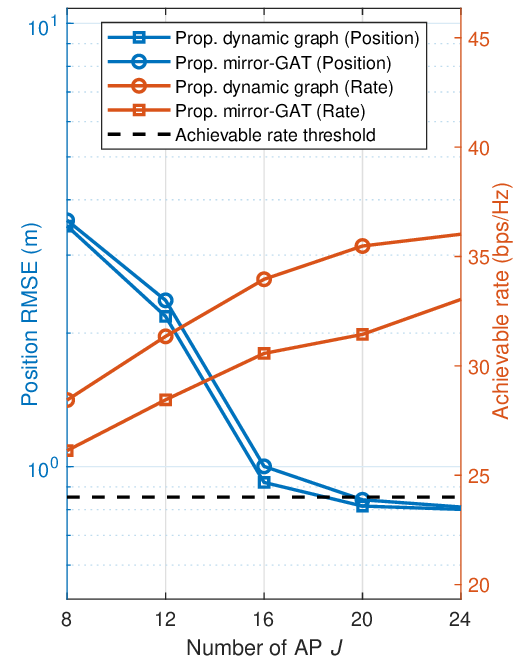}}
\subfloat[Velocity RMSE]{
		\includegraphics[width=1.75 in,scale=0.5]{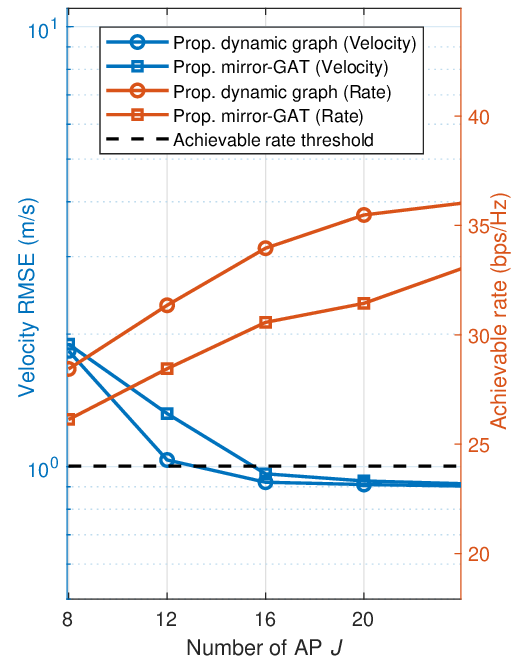}}
\caption{Sensing performance versus the number of AP $J$.}
\label{Sim:RMSE_AP}
\vspace{-0.5cm}
\end{figure}

\textcolor{black}{Fig.~\ref{Sim:RMSE_AP} illustrates how AP deployment density (number of APs) affects sensing and communication performance. Both proposed frameworks consistently meet the predefined communication sum-rate requirements across all AP densities. Increasing the number of APs progressively improves position and velocity estimation accuracy and simultaneously enhances achievable communication rates, confirming that denser AP deployment substantially benefits cooperative cell-free ISAC networks in terms of sensing precision and communication quality.}

\subsection{Trade-off Analysis}
\begin{figure}[!t]
\centering
\subfloat[Position RMSE]{
		\includegraphics[width=1.7 in,scale=0.5]{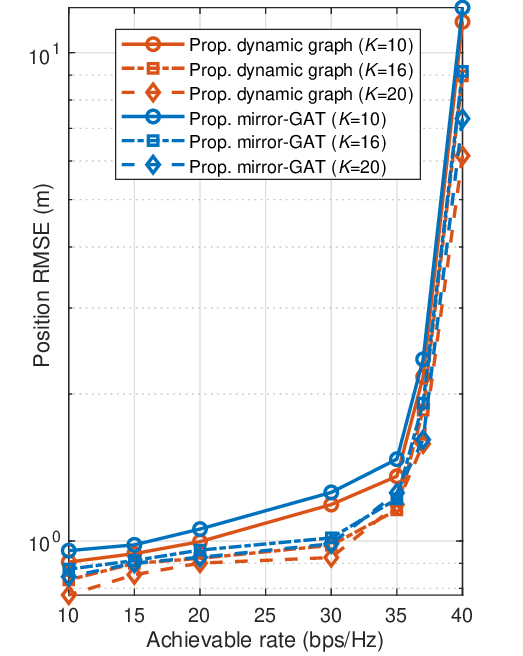}}
\subfloat[Velocity RMSE]{
		\includegraphics[width=1.7 in,scale=0.5]{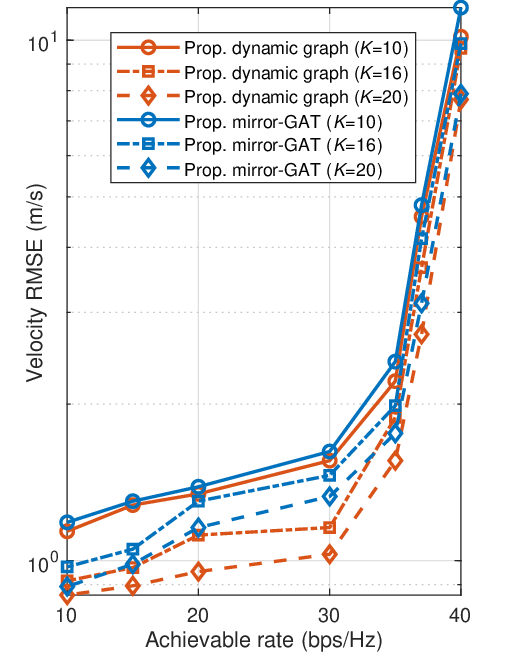}}
\caption{Trade off analysis under different $K$.}
\label{Sim:trade_off_k} 
\vspace{-0.3 cm}
\end{figure}

\begin{figure}[!t]
\centering
\subfloat[Position RMSE]{
		\includegraphics[width=1.7 in,scale=0.5]{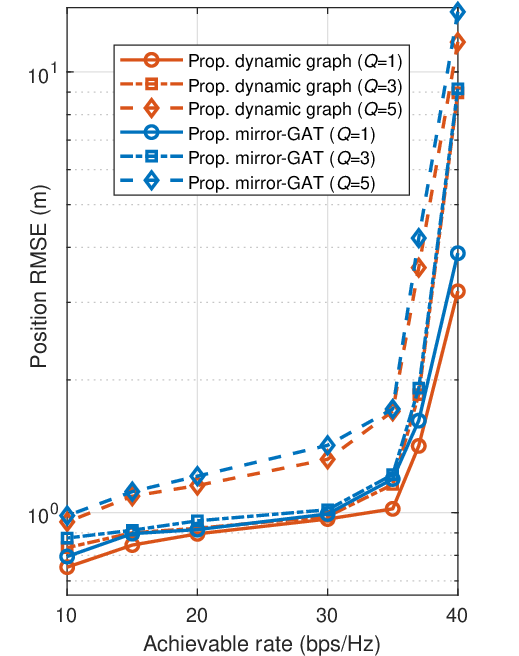}}
\subfloat[Velocity RMSE]{
		\includegraphics[width=1.7 in,scale=0.5]{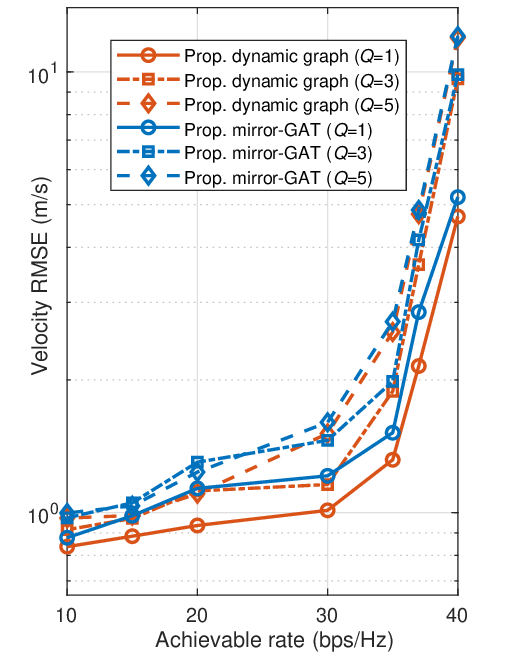}}
\caption{Trade off analysis under different $Q$.}
\label{Sim:trade_off_q}
\vspace{-0.0 cm}
\end{figure}

\textcolor{black}{Figs.~\ref{Sim:trade_off_k} and \ref{Sim:trade_off_q} illustrate the trade-off between multi-user communication and sensing performance, characterized by the achievable communication sum-rate threshold $\gamma$ and the corresponding position and velocity RMSEs. Clearly, imposing higher sum-rate constraints reduces resources available for sensing, thereby degrading estimation accuracy and increasing both position and velocity RMSE values. Specifically, Fig.~\ref{Sim:trade_off_k} shows the impact of the number of CUs ($K$). Increasing $K$ slightly improves sensing accuracy at a given sum-rate constraint, since serving more users provides additional spatial diversity. Consequently, communication sum-rate requirements can be satisfied more efficiently, leaving more resources available for target illumination and sensing, thus reducing position and velocity RMSEs. Conversely, Fig.~\ref{Sim:trade_off_q} demonstrates that increasing the number of sensing targets ($Q$) deteriorates estimation performance. As more targets compete for limited sensing resources, the illumination energy allocated to each individual target inevitably decreases. Additionally, for a fixed compressed‑feature dimension, increasing the number of targets results in greater feature information loss during data fusion, further negatively impacting estimation accuracy. These observations highlight the critical importance of carefully balancing resource allocation between communication and sensing functionalities, and clearly demonstrate that our proposed graph-learning-based frameworks effectively manage this trade-off to achieve superior performance in cooperative cell-free ISAC networks.}

\begin{figure}[!t]
\centering
\includegraphics[width=3.4 in]{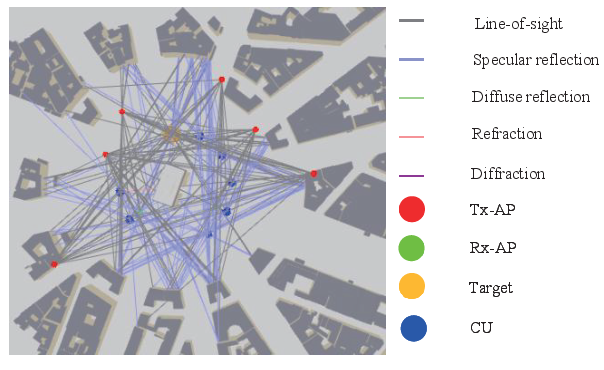}
\caption{Multipath propagation scenario of cell-free ISAC using ray-tracing Sionna-RT simulator.}
\label{Sim:sionna} \vspace{-0.1 cm}
\end{figure}

\subsection{Practical Analysis}
\textcolor{black}{To further examine the practical feasibility of the proposed algorithms, we evaluate their performance under two key factors commonly encountered in real-world ISAC deployments. First, conventional Rayleigh and Rician channel models may inadequately represent the intricate propagation characteristics of urban ISAC environments, characterized by intense multipath and clutter effects. To obtain a more realistic assessment, we utilize the ray-tracing-based NVIDIA Sionna‑RT toolbox \cite{Sionna} to generate physically accurate propagation scenarios of a cell-free ISAC system, as illustrated in Fig.~\ref{Sim:sionna}. The simulation incorporates key propagation phenomena such as specular and diffuse reflections, refraction, and diffraction, effectively capturing typical urban multipath and clutter-rich conditions.
Second, imperfect channel estimation is another critical factor affecting ISAC performance in practice. To assess the robustness of the proposed algorithms under this impairment, we model the estimated CSI as \be \widehat{\mathbf{h}}_{j,i,k} = \mathbf{h}_{j,i,k} +\Delta\mathbf{h}, \,\, \Delta\mathbf{h}\sim\mathcal{CN}(0, \sigma_{\rm e}^2\mathbf{I}_M), \ee  where $\sigma_{\mathrm e}^2$ denotes the channel estimation error variance. The corresponding results are summarized in Table~\ref{tab:csi_results}, where gray rows represent the dynamic graph learning framework and white rows represent the mirror‑GAT framework. As observed, both proposed methods maintain robust sensing accuracy while meeting the communication sum-rate requirements under practical impairments. Furthermore, the dynamic graph learning approach demonstrates stronger generalization capability in complex and heterogeneous environments, whereas the mirror‑GAT framework achieves comparable robustness with considerably lower computational complexity and signaling overhead.}

\begin{table}[!t]
  \centering
  \footnotesize
  \caption{Practical analysis under various non-ideal scenarios. ($J=8$, $K=8$, $Q=1$, $\gamma=$10bps/Hz)}
  \label{tab:csi_results}
  \begin{tabular}{ c  S[table-format=2.2] S[table-format=2.2] S[table-format=2.2] }
    \hline
     Metric & {Ideal} & {Sionna-RT} & {Imperfect}   \\
    \hline
    \multirow{2}{*}{Position RMSE (m)} & \cellcolor{gray!15}1.53 & \cellcolor{gray!15}1.64 & \cellcolor{gray!15}1.55 \\
                                                   & 1.61 & 1.69 & 1.69 \\
    \cmidrule(lr){1-4}
    \multirow{2}{*}{Velocity RMSE (m/s)} & \cellcolor{gray!15}1.29 & \cellcolor{gray!15}1.32  & \cellcolor{gray!15}1.33 \\
                                                   &1.47 & 1.551 & 1.98 \\
    \cmidrule(lr){1-4}
    \multirow{2}{*}{Achievable Rate (bps/Hz)} & \cellcolor{gray!15}17.65 & \cellcolor{gray!15}16.56 & \cellcolor{gray!15}15.61 \\
                                                     & 15.14 & 14.07 & 12.75 \\
    \hline
  \end{tabular} \vspace{-0.3 cm}
\end{table}

\begin{figure}[!t]
\centering
\subfloat[Compressed‑feature dimension $\varrho$]{
		\includegraphics[width=1.7 in,scale=0.5]{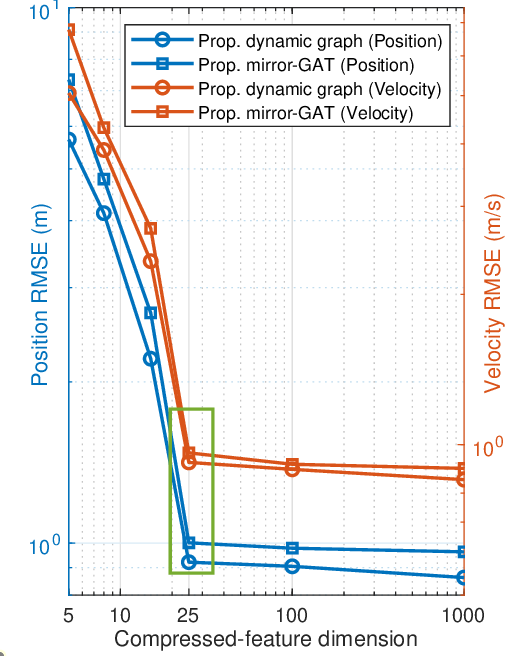}}
\subfloat[Snapshot~$\&$~Iteration]{
		\includegraphics[width=1.7 in,scale=0.5]{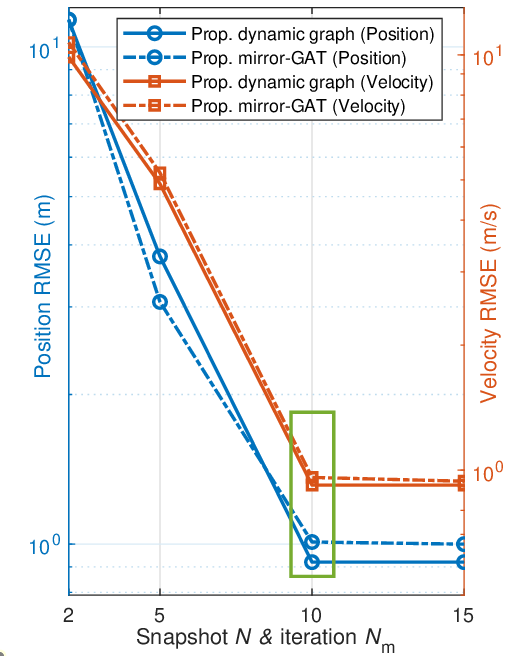}}
\caption{Sensing performance versus key parameters of learning frameworks.}
\label{Sim:frame_para}
\vspace{-0.5cm}
\end{figure}

\textcolor{black}{Fig.~\ref{Sim:frame_para} investigates the impact of key parameters within the learning frameworks on sensing performance. Specifically, Fig.~\ref{Sim:frame_para}(a) evaluates how the compressed‑feature dimension $\varrho$ in the 3D-CNN module affects position and velocity estimation accuracy. As $\varrho$ increases, both proposed frameworks show significant improvements in estimation performance. This is primarily because a larger compressed‑feature dimension preserves richer feature information from the echo signals. Nevertheless, an increased $\varrho$ leads to greater backhaul overhead due to the transmission of more comprehensive feature data. At a moderate compressed‑feature dimension ($\varrho = 25$), the performance degradation remains minimal compared to higher scales, yet considerably reduces signaling overhead.
Fig.~\ref{Sim:frame_para}(b) demonstrates the relationship between the number of snapshots/iterations and the accuracy of the proposed graph learning schemes. Increasing these parameters enhances estimation performance by enabling more extensive multi-perspective observations and iterative refinements.  Nonetheless, this advantage comes with increased computational complexity, overhead, and processing latency. Notably, estimation accuracy stabilizes beyond $N/N_{\mathrm{m}} = 10$, indicating diminishing returns from further increments. Consequently, setting $\varrho = 25$ and $N/N_{\mathrm{m}} = 10$ provides an optimal balance, which is employed consistently in prior simulations and underscores their suitability for practical large-scale cell-free ISAC deployments.}

\subsection{Computational Complexity}
\begin{figure}[!t]
\centering
\subfloat[Model parameter number]{
		\includegraphics[width=1.7 in,scale=0.5]{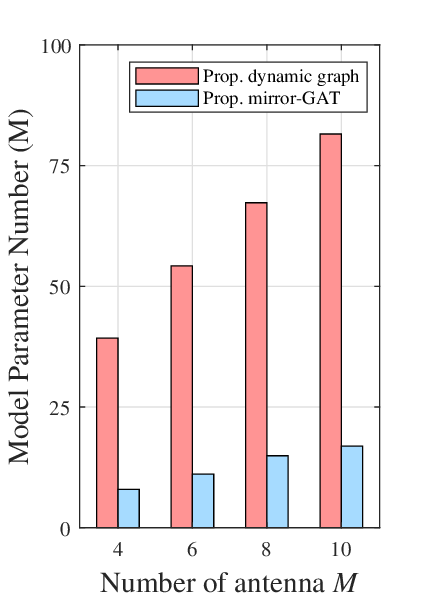}}
\subfloat[Average execution time]{
		\includegraphics[width=1.7 in,scale=0.5]{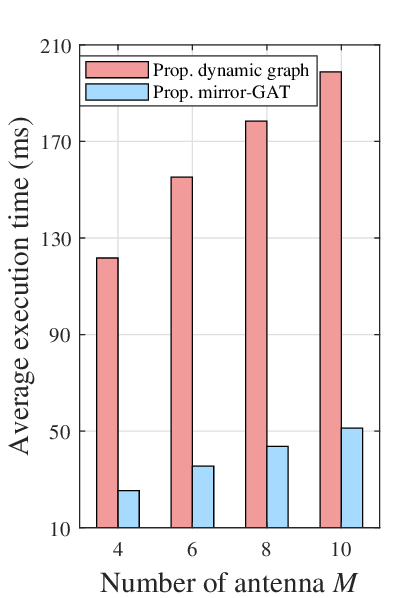}}
\caption{Complexity analysis.}
\vspace{-0.5 cm}
\label{Sim:complexity}
\end{figure}

To evaluate the practical feasibility of the proposed graph-learning frameworks, we analyze their computational complexity in terms of model size (number of parameters) and average execution time under different numbers of transmit antennas ($M$). As antenna count significantly influences computational demands at each AP, it serves as a key indicator for practical implementation complexity. The reported execution times in Fig.~\ref{Sim:complexity} include both local inference at APs and the simulated backhaul signaling overhead, modeled as an additional delay proportional to the transmitted data size.
As shown in Fig.~\ref{Sim:complexity}, the mirror-GAT framework consistently demonstrates lower computational complexity, characterized by significantly fewer model parameters and reduced execution time compared to the dynamic graph learning framework. These results confirm that the mirror-GAT approach is highly suitable for practical, large-scale deployments, as it effectively balances computational efficiency and signaling overhead while maintaining competitive performance in multi-user communication and target sensing accuracy.

\section{Conclusions}
\label{Conclusions}
In this paper, we investigated joint network configuration and multi-target parameter estimation in cooperative cell-free ISAC systems. We proposed a dynamic graph learning framework employing structural and temporal dual-attention mechanisms, as well as a lightweight mirror-GAT framework that significantly reduces computational complexity and signaling overhead. Extensive simulations confirmed that both frameworks substantially outperform heuristic and recent deep-learning-based schemes, achieving superior multi-user communication performance and robust, high-precision sensing accuracy (position RMSE $\! < \! 1$m, velocity RMSE $ \! < \!1$m/s) across various settings. Notably, mirror-GAT offers comparable performance with significantly reduced complexity, underscoring its practical feasibility for large-scale cell-free ISAC deployments.

\end{document}